\documentclass[aps,twocolumn,superscriptaddress,amsmath,amssymb,amsfonts]%
{revtex4-2}

\usepackage{nicefrac}

\usepackage{graphicx}
\usepackage[caption=false]{subfig}
\usepackage{ragged2e}
\DeclareCaptionJustification{justified}{\justifying}
\usepackage{float}
\floatname{circuit}{\small CIRCUIT~}

\usepackage{braket}
\usepackage{qcircuit}
\usepackage{environ}
\newcommand{\myqctmp}[2][0.25]{\Qcircuit @C=#2em @R=#1em @!R}
\NewEnviron{myqcircuit}[1][0.25]{\vcenter{\myqctmp[#1]{0.75} {\BODY}}}
\NewEnviron{myqcircuit*}[2]{\vcenter{\myqctmp[#1]{#2} {\BODY}}}

\newcommand{\Ry}{R_y}
\newcommand{\Rz}{R_z}
\newcommand{\CNOT}{\text{CNOT}}
\newcommand{\Uprod}{U_{\text{prod}}}
\newcommand{\Usum}{U_{\text{sum}}}
\newcommand{\Usumsum}{U_{\text{sum2}}}

\usepackage[colorlinks,urlcolor=blue,linkcolor=blue,citecolor=blue,breaklinks=true]{hyperref}
\usepackage[capitalize,nameinlink]{cleveref}
\crefname{circuit}{Circuit}{Circuits}
\crefformat{subcircuit}{#2Circuit~\thecircuit#1#3}
\crefrangeformat{subcircuit}{#3Circuits~\thecircuit#1#4 to~#5\thecircuit#2#6}
\crefmultiformat{subcircuit}{#2Circuits~\thecircuit#1#3}%
{ and~#2\thecircuit#1#3}{, #2\thecircuit#1#3}{ and~#2\thecircuit#1#3}

\newcommand{\ehtabs}{\textbf{\textbf{EHands}}}
\newcommand{\eh}{\textbf{\texttt{EHands}}}
\newcommand{\ehp}{\textbf{\texttt{EHands protocol}}}
\newcommand{\ehptabs}{\textbf{\textbf{EHands protocol}}}

\newcommand{\qcrank}{\textbf{\texttt{QCrank}}}

\begin{document}

\title{EHands: Quantum Protocol for Polynomial Computation on Real-Valued Encoded States}

\author{Jan Balewski}
\affiliation{National Energy Research Scientific Computing Center,
             Lawrence Berkeley National Laboratory,
             Berkeley, CA 94720, USA}

\author{Chris~Pestano}
\affiliation{Computer Science Department,
            San Francisco State University,
            San Francisco, CA 94132, USA}

\author{Mercy G. Amankwah}
\affiliation{Department of Mathematics, Applied Mathematics and Statistics,
             Case Western Reserve University,
             Cleveland, OH 44106, USA}

\author{E. Wes Bethel}
\affiliation{Computer Science Department,
             San Francisco State University,
             San Francisco, CA 94132, USA}

\author{Talita Perciano}
\email{Corresponding authors: tperciano@lbl.gov, rvanbeeumen@lbl.gov}
\affiliation{Scientific Data Division,
             Lawrence Berkeley National Laboratory,
             Berkeley, CA 94720, USA}

\author{Roel Van Beeumen}
\email{Corresponding authors: tperciano@lbl.gov, rvanbeeumen@lbl.gov}
\affiliation{Applied Mathematics and Computational Research Division,
             Lawrence Berkeley National Laboratory,
             Berkeley, CA 94720, USA}

\date{\today}

\begin{abstract}
We present \eh, a \textit{quantum-native protocol} for implementing multivariable polynomial transformations on quantum processors. The protocol introduces four fundamental, reversible operators—\textit{multiplication}, \textit{addition}, \textit{negation}, and \textit{parity flip}—and employs the Expectation Value ENcoding (EVEN) scheme to represent real numbers as quantum states. Unlike discretization or binary encoding methods, \eh\ operates directly on vectorized real-valued inputs prepared in the initial state and applies a shallow quantum circuit that depends only on the polynomial coefficients. The result is obtained from the expectation value measured on a single qubit, enabling efficient parallel evaluation of a polynomial across multiple data points using a single circuit. We introduce both a \textit{reversible} implementation for degree-$d$ polynomials, requiring $3d$ qubits, and a \textit{non-reversible} variant that uses qubit resets to reduce the requirements to $d+1$ qubits. Both implementations exhibit linear depth scaling in $d$ and are explicitly decomposed into one- and two-qubit gates for direct execution on current quantum processing units. The protocol's effectiveness is demonstrated through experimental validation on IBM's Heron-class quantum processors, showing reliable polynomial approximations of functions like ReLU and arctan.
\end{abstract}


\maketitle

\section{Introduction}

Quantum computing promises to accelerate computation across a wide range of scientific disciplines, with significant breakthroughs such as Shor's factoring algorithm~\cite{shor1994}, Grover's search method~\cite{grov1996}, and the HHL algorithm for solving linear systems~\cite{haha2009}.
While problems like quantum dynamics simulation~\cite{low2017optimal} naturally map to quantum hardware, classical computers remain superior for arithmetic operations. 
Efficient quantum implementations of addition, multiplication, and polynomial evaluation are crucial for enabling end-to-end quantum workflows. It allows avoiding repeated measurement, costly classical post-processing, and  quantum state reinitializations, that hinder hybrid approaches.

Polynomial transformations play a key role in interpolation, approximation, and modeling of non-linear functions across science and engineering.
Yet current quantum algorithms for such transformations remain limited: they often mimic classical  algorithms based on discretization or rely on binary encoding~\cite{veba1996,taku2005,taku2008,taka2009,wali2025}, leading to deep circuits or large ancilla overhead, impractical for near-term devices.
Alternative methods such as quantum signal processing (QSP)~\cite{loch2017,doli2024}, quantum singular value transformation (QSVT)~\cite{gisu2019}, and quantum eigenvalue transformation (QET)~\cite{doli2022} enable implicit polynomial transformations of singular values or eigenvalues, but are not designed for direct polynomial evaluation on input data prepared in initial quantum states.

Quantum algorithms also differ widely in their sampling requirements. Grover's algorithm and quantum phase estimation (QPE) are nearly deterministic, requiring only tens of repetitions~\cite{grover-shots,phase-estimation-shots}, whereas algorithms like quantum approximate optimization algorithm (QAOA), quantum Monte Carlo, and boson sampling demand thousands to millions of shots to yield reasonable accuracy~\cite{qaoa-shots,qmc-shots,boson-shots}. The protocol introduced here, \eh, falls into the latter category: it encodes results in expectation values and thus requires sampling, but achieves this through shallow, structured circuits.

Hybrid variational methods~\cite{recioarmengol2025polynomial} can approximate functions via parameter optimization but require extensive classical-quantum feedback, which is resource-intensive and noise-sensitive. In contrast, \eh\ provides a direct, deterministic quantum-native approach to polynomial evaluation.

This paper makes four main contributions. 
First, we introduce \eh, a protocol that implements a universal set of basic arithmetic operations directly on QPUs using shallow, reversible circuits. Building on these primitives, \eh\ enables multivariable polynomial transformations on data encoded via  the Expectation Value ENcoding (EVEN) scheme.
Second, \eh\ naturally operates on vectorized inputs, allowing efficient parallel  polynomial evaluation across multiple data points in a single quantum circuit. 
Third, we present both \emph{reversible} and \emph{non-reversible} implementations for degree-$d$ polynomials. The reversible version requires $3d$ qubits, while the non-reversible variant leverages resets to reduce the qubit requirement to $d + 1$. Both implementations achieve linear depth scaling in $d$ and are explicitly decomposed in one- and two-qubit gates for direct execution on current QPUs.
Four, we demonstrate the practicality of \eh\ through experiments on contemporary NISQ hardware, showing reliable accuracy for polynomial approximations of common nonlinear functions.

\section{Expectation-value encoding (EVEN) scheme}

To encode a list of $N \geq 1$ real values $x_i \in [-1,1]$, we construct a product state $\ket{\phi}$ with $N$ qubits, each initialized to the $\ket{0}$ state and subjected to an $\Ry(\theta_i)$ rotation, where $\theta_i = \arccos(x_i)$.

To retrieve data from a specific qubit in the EVEN state $\ket{\phi}$, we apply \emph{decoding} by measuring the expectation value (EV) of the Pauli-$Z$ operator for that qubit, given by $\hat{x} = \bra{\phi}O\ket{\phi}$.
\Cref{fig:eh-EVEN} shows the encoding and decoding of an EVEN state for a single real input $x$.
See \cref{secSM:encode-decode} for encoding and decoding of 2 real numbers on 2 qubits.

The EVEN is a type of angle encoding~\cite{rada2024}. Rather than using a simple linear transformation to convert a real number into a rotation angle within the range $[0,\pi]$, EVEN applies a non-linear transformation via the $\arccos$ function. This seemingly minor modification plays a crucial role in simplifying the polynomial transformation protocol described below.

\section{The \ehptabs}

We introduce four fundamental reversible quantum operations: (1)~\emph{product-with-memory}, (2)~\emph{weighted sum}, (3)~\emph{negation}, and (4)~\emph{parity flip}. 
\Cref{fig:eh-prod,fig:eh-sum,fig:eh-sign,fig:eh-flip} present their corresponding quantum circuits.
These operations  will serve as the building blocks for polynomial calculations.

The {\bf product-with-memory} operation is defined by the unitary $\Uprod = \CNOT_{0,1} \cdot \left[ I \otimes \Rz(\nicefrac{\pi}{2}) \right]$. \Cref{fig:eh-prod} shows its implementation. For the pair of inputs $x_0, x_1$, the information about their product is stored in qubit $q_1$, while qubit $q_0$ retains the value of $x_0$. The value of $x_0 x_1$ can be retrieved by measuring the EV of the operator $O_p = I \otimes \sigma_z$. Alternatively, additional quantum operations can be applied to either qubit for further computations.

The {\bf weighted sum} operation is defined by $\Usum(w) = \left[ \Ry(\nicefrac{-\alpha}{2}) \otimes I \right] \cdot \CNOT_{1,0} \cdot \left[ \Ry(\nicefrac{\alpha}{2}) \otimes I \right] \cdot \Uprod$, shown in \cref{fig:eh-sum}. The parameter $\alpha = \arccos(1 - 2w)$ depends on the chosen relative weight $w\in[0,1]$, making $\Usum(w)$ a parametric unitary operator. It computes the weighted sum $S_w(x_0,x_1) = w x_0 + (1 - w) x_1$, with the result stored in qubit $q_0$. The value of $S_w(x_0,x_1)$ can be retrieved by measuring the EV of the complementary operator $O_s = \sigma_z \otimes I$. Again, further arithmetic operations can be performed on the data encoded in qubit $q_0$. Strictly speaking, the weighted sum circuit could also be used for multiplication since $q_1$ encodes the product $x_0 x_1$. However, in practice, we prefer the dedicated multiplication circuit shown in \cref{fig:eh-prod}, which requires only 1 CNOT gate. 
Although qubit $q_1$ in \cref{fig:eh-sum} is no longer used it must be preserved for reversibility.

\begin{figure}[t]
\captionsetup[subfloat]{position=top}
\centering
\subfloat[\raggedright EVEN encoding and decoding\label{fig:eh-EVEN}]{%
\begin{minipage}{\columnwidth}
\begin{flushleft}
\scalebox{0.9}{$\hspace{1.75em}\begin{myqcircuit}
& \lstick{\ket{0}} & \gate{\Ry(\theta)} & \rstick{\ket{\phi}}\qw
\end{myqcircuit}%
\qquad\quad\longrightarrow\qquad\ %
\begin{myqcircuit}
& \lstick{\ket{\phi}} & \measuretab{Z} & \rstick{\hat{y} = \text{EV}(\sigma_z) \simeq x}
\end{myqcircuit}$}%
\end{flushleft}
\end{minipage}}

\subfloat[\raggedright product-with-memory (multiplication)\label{fig:eh-prod}]{%
\begin{minipage}{\columnwidth}
\begin{flushleft}
\scalebox{0.9}{$\hspace{1.75em}\begin{myqcircuit}
& \lstick{\ket{0}_0} & \gate{\Ry(\theta_0)} & \qw & \qw & \ctrl{1} & \qw & \qw \\
& \lstick{\ket{0}_1} & \gate{\Ry(\theta_1)} & \qw & \gate{\Rz(\frac{\pi}{2})} & \targ & \qw & \measuretab{Z} & \rstick{\hat{y}_1 = \text{EV}(I \otimes \sigma_z) \simeq x_0 x_1}
{\gategroup{1}{5}{2}{6}{1.2em}{--}}
\end{myqcircuit}$}%
\end{flushleft}
\end{minipage}}

\subfloat[\raggedright weighted sum (addition) \hfill $\hat{y}_0 = \text{EV}(\sigma_z \otimes I)\simeq w x_0 + (1 - w) x_1$\label{fig:eh-sum}]{%
\begin{minipage}{\columnwidth}
\begin{flushleft}
\scalebox{0.9}{$\hspace{1.75em}\begin{myqcircuit}
& \lstick{\ket{0}_0} & \gate{\Ry(\theta_0)} & \qw & \qw & \ctrl{1} & \gate{\Ry(\frac{\alpha}{2})} & \targ & \gate{\Ry(-\frac{\alpha}{2})} & \qw & \measuretab{Z} & \rstick{\hat{y}_0} \\
& \lstick{\ket{0}_1} & \gate{\Ry(\theta_1)} & \qw & \gate{\Rz(\frac{\pi}{2})} & \targ & \qw & \ctrl{-1} & \qw & \qw & \qw
{\gategroup{1}{5}{2}{9}{1.2em}{--}}
\end{myqcircuit}$}%
\end{flushleft}
\end{minipage}}

\subfloat[\raggedright sign change (negation)\label{fig:eh-sign}]{%
\begin{minipage}{0.625\columnwidth}
\begin{flushleft}
\scalebox{0.9}{$\hspace{1.75em}\begin{myqcircuit}
& \lstick{\ket{0}} & \gate{\Ry(\theta)} & \qw & \gate{X} & \qw & \measuretab{Z} 
{\gategroup{1}{5}{1}{5}{1.2em}{--}}\\
&&&&& \rstick{\hat{y} \simeq -x}
\end{myqcircuit}$}%
\end{flushleft}
\end{minipage}}%
\subfloat[\raggedright parity flip\label{fig:eh-flip}]{%
\begin{minipage}{0.375\columnwidth}
\begin{flushleft}
\scalebox{0.9}{$\hspace{0.5em}\begin{myqcircuit}[0.75]
& \qw & \qw & \qw & \qw & \qw & \qw & \ctrl{1} & \qw & \qw \\
&&&& \lstick{\ket{0}_{\text{anc}}} & \qw & \gate{H} & \ctrl{0} & \qw & \qw\\
{\gategroup{1}{7}{2}{8}{1.2em}{--}}
\end{myqcircuit}$}%
\end{flushleft}
\end{minipage}}%
\vspace{-2ex}
\caption{%
(a) EVEN encoding of the real number \( x \) into the quantum state \( \ket{\phi} \) using $\theta=\arccos(x)$, along with its decoding, where the expectation value of the Pauli-\( Z \) operator is the estimator \( \hat{x} \).
The four building blocks of the \ehp\ are represented as circuits : (b) multiplication, (c) addition, (d) negation, and (e) parity flip.
\label{fig:eh-3blocks}}
\end{figure}   

The {\bf negation} operation is implemented using the standard $X$-gate, as shown in \cref{fig:eh-sign}. 
Transforming quantum state $x \to -x$ is particularly useful for overcoming the constraint that the weight $w$ of the $S_w(x_0,x_1)$  operation must be positive. By applying negation to one of the inputs, such as on qubit $q_1$ in \cref{fig:eh-sum}, the result becomes a weighted difference $S_w(x_0,-x_1) = w x_0 - (1 - w) x_1$.

Finally, {\bf parity flip} operation, implemented by the circuit shown in \cref{fig:eh-flip}, is required when multiple addition operations are performed sequentially, e.g., in constructing a polynomial in \cref{fig:eh-poly_4-rev}. 
The weighted sum operations can introduce phase factors that manifest as imaginary components in intermediate calculations. 
The parity flip operation uses an ancilla qubit in superposition to apply controlled-$Z$ gates that cancel these unwanted phase contributions, ensuring the final result remains real-valued as required for polynomial evaluation.

We refer to \cref{secSM:EHands} for unitary representations and proofs of the correctness of arithmetic and concatenation operations for those circuits.

\begin{figure*}[htbp]
    \captionsetup[subfloat]{position=top}
    \centering
    \subfloat[Reversible \eh\ circuit\label{fig:eh-poly_4-rev}]{%
    \scalebox{0.9}{$\begin{myqcircuit}
    \\ 
    & \lstick{\ket{0}_{a_0}} & \gate{\Ry(\varphi_0)} & \qw \barrier[-1.75em]{11} & \qw & \qw & \qw & \qw & \ustick{\textcolor{blue}{a_0}} \qw & \qw \barrier[-.25em]{11} & \qw & \qw & \qw & \qw & \qw & \qw & \qw \barrier[-.5em]{11} & \qw & \qw & \qw & \qw & \qw & \qw & \qw & \qw & \qw & \qw & \qw & \qw & \qw & \multigate{2}{\Sigma\frac{1}{5}} & \qw \barrier[-1.5em]{11} & \measuretab{Z} 
    \\ 
    & \lstick{\ket{0}_{d_0}} & \gate{\Ry(\theta)} & \qw & \multigate{2}{\Pi}_<<{0} & \qw & \qw & \qw & \ustick{\textcolor{blue}{x}} \qw & \qw & \qw & \multigate{1}{\Pi} & \qw   & & & & && &     & & & & && &     & & & & & \rstick{\textcolor{blue}{P_4(x)}}
    \\ 
    & \lstick{\ket{0}_{a_1}} & \gate{\Ry(\varphi_1)} & \qw & \ghost{\Pi} & \qw & \qw & \qw & \ustick{\textcolor{blue}{a_1}} \qw & \qw & \qw & \ghost{\Pi} & \qw & \qw & \ustick{\textcolor{blue}{a_1x\phantom{^1}}} \qw & \qw & \qw & \qw & \qw & \qw & \qw & \qw & \qw & \qw &  \qw & \qw & \multigate{2}{\Sigma\frac{1}{4}} & \qw & \ctrl{9} & \qw & \ghost{\Sigma\frac{1}{5}} & &
    \\ 
    & \lstick{\ket{0}_{d_1}} & \gate{\Ry(\theta)} & \qw & \ghost{\Pi}_<<{1} & \multigate{2}{\Pi}_<<{0} & \qw & \qw & \ustick{\textcolor{blue}{x^2}} \qw & \qw & \qw & \multigate{1}{\Pi} & \qw 
    \\ 
    & \lstick{\ket{0}_{a_2}} & \gate{\Ry(\varphi_2)} & \qw & \qw & \ghost{\Pi} & \qw & \qw & \ustick{\textcolor{blue}{a_2}} \qw & \qw & \qw & \ghost{\Pi} & \qw & \qw & \ustick{\textcolor{blue}{a_2x^2}} \qw & \qw & \qw & \qw & \qw & \qw & \qw &  \qw & \multigate{2}{\Sigma\frac{1}{3}} & \qw & \ctrl{6} & \qw & \ghost{\Sigma\frac{1}{4}} & \qw &
    \\ 
    & \lstick{\ket{0}_{d_2}} & \gate{\Ry(\theta)} & \qw & \qw & \ghost{\Pi}_<<{1} & \multigate{2}{\Pi}_<<{0} & \qw & \ustick{\textcolor{blue}{x^3}} \qw & \qw & \qw & \multigate{1}{\Pi} & \qw 
    \\ 
    & \lstick{\ket{0}_{a_3}} & \gate{\Ry(\varphi_3)} & \qw & \qw & \qw & \ghost{\Pi} & \qw & \ustick{\textcolor{blue}{a_3}} \qw & \qw & \qw & \ghost{\Pi} & \qw & \qw & \ustick{\textcolor{blue}{a_3x^3}} \qw & \qw & \qw & \qw & \multigate{2}{\Sigma\frac{1}{2}} & \qw & \ctrl{3} & \qw & \ghost{\Sigma\frac{1}{3}} & \qw & & &
    \\ 
    & \lstick{\ket{0}_{d_3}} & \gate{\Ry(\theta)} & \qw & \qw & \qw & \ghost{\Pi}_<<{1} & \qw & \ustick{\textcolor{blue}{x^4}} \qw & \qw & \qw & \multigate{1}{\Pi} & \qw 
    \\ 
    & \lstick{\ket{0}_{a_4}} & \gate{\Ry(\varphi_4)} & \qw & \qw & \qw & \qw & \qw & \ustick{\textcolor{blue}{a_4}} \qw & \qw & \qw & \ghost{\Pi} & \qw & \qw & \ustick{\textcolor{blue}{a_4x^4}} \qw & \qw & \qw & \qw & \ghost{\Sigma\frac{1}{2}} & \qw 
    \\ 
    & \lstick{\ket{0}_{\text{anc}_0}} & \qw & \qw & \qw & \qw & \qw & \qw & \qw & \qw & \qw & \qw & \qw & \qw & \qw & \qw & \qw & \qw & \qw & \gate{H} & \control \qw & \qw 
    \\ 
    & \lstick{\ket{0}_{\text{anc}_1}} & \qw & \qw & \qw & \qw & \qw & \qw & \qw & \qw & \qw & \qw & \qw & \qw & \qw & \qw & \qw & \qw & \qw & \qw & \qw &  \qw & \qw & \gate{H} & \control \qw & \qw 
    \\ 
    & \lstick{\ket{0}_{\text{anc}_2}} & \qw & \qw & \qw & \qw & \qw & \qw & \qw & \qw & \qw & \qw & \qw & \qw & \qw & \qw & \qw & \qw & \qw & \qw & \qw &  \qw & \qw & \qw & \qw & \qw & \qw & \gate{H} & \control \qw & \qw & & &\\
    \\ 
    \end{myqcircuit}$}
    } \\
    \subfloat[Non-reversible \eh\ circuit\label{fig:eh-poly_4-irrev}]{%
    \hspace{2.125em}\scalebox{0.9}{$\begin{myqcircuit*}{0.25}{0.375}
    & \lstick{\ket0_{d_0}\!} & \gate{\Ry(\theta)} & \qw \barrier[-1.25em]{4} & \multigate{1}{\Pi} & \qw & \qw & \qw & \ustick{\textcolor{blue}{x}} \qw & \qw & \qw & \qw & \qw & \qw & \qw & \qw & \qw & \qw & \qw & \qw & \qw & \qw & \qw & \qw & \qw & \multigate{1}{\Pi} & \qw & \push{\ket{0}} & \gate{\Ry(\varphi_0)} & \qw & \ustick{\textcolor{blue}{a_0~}} \qw & \qw & \qw & \multigate{1}{\Sigma{\frac{1}{5}}} & \qw \barrier[-1.25em]{4} & \measuretab{Z} \\
    & \lstick{\ket0_{d_1}\!} & \gate{\Ry(\theta)} & \qw & \ghost{\Pi} & \multigate{1}{\Pi} & \qw & \qw & \ustick{\textcolor{blue}{x^2}} \qw & \qw & \qw & \qw & \qw & \qw & \qw  & \qw & \qw & \qw & \qw & \multigate{1}{\Pi} & \qw & \push{\ket{0}} & \gate{\Ry(\varphi_1)} & \qw & \qw & \ghost{\Pi} & \qw & \ustick{\textcolor{blue}{a_1x~\,}} \qw & \multigate{1}{\Sigma{\frac{1}{4}}} & \qw & \qw & \qw & \ctrl{1} & \ghost{\Sigma{\frac{1}{5}}} \\
    & \lstick{\ket0_{d_2}\!} & \gate{\Ry(\theta)} & \qw & \qw & \ghost{\Pi} & \multigate{1}{\Pi} & \qw & \ustick{\textcolor{blue}{x^3}} \qw & \qw & \qw & \qw & \qw & \multigate{1}{\Pi} & \qw & \push{\ket{0}} & \gate{\Ry(\varphi_2)} & \qw & \qw & \ghost{\Pi} & \qw & \ustick{\textcolor{blue}{a_2x^2~}} \qw & \multigate{1}{\Sigma{\frac{1}{3}}} & \qw & \qw & \qw & \ctrl{1} & \qw & \ghost{\Sigma{\frac{1}{4}}} & \qw & \push{\ket{0}} & \gate{H} & \control \qw & \push{\hspace{3ex}} \qw \\ 
    & \lstick{\ket0_{d_3}\!} & \gate{\Ry(\theta)} & \qw & \qw & \qw & \ghost{\Pi} & \qw & \ustick{\textcolor{blue}{x^4}} \qw & \multigate{1}{\Pi} & \qw & \push{\ket{0}} & \gate{\Ry(\varphi_3)} & \ghost{\Pi} & \qw & \ustick{\textcolor{blue}{a_3x^3~}} \qw & \multigate{1}{\Sigma{\frac{1}{2}}} & \qw & \qw & \qw & \ctrl{1} & \qw & \ghost{\Sigma{\frac{1}{3}}} & \qw & \push{\ket{0}} & \gate{H} & \control \qw & \qw \\
    & \lstick{\ket{0}_{\text{anc}}\!} & \gate{\Ry(\varphi_4)} & \qw & \qw & \qw & \qw & \qw & \qw & \ghost{\Pi} & \qw & \ustick{\textcolor{blue}{\raisebox{-2ex}{\text{$a_4x^4~$}}}} \qw & \qw & \qw & \qw & \qw & \ghost{\Sigma{\frac{1}{2}}} & \qw & \push{\ket{0}} & \gate{H} & \control \qw & \qw & & & & & & & & & & & & & & \\
    && \uparrow &&&&&&&&&& \uparrow &&&& \uparrow &&&&&& \uparrow &&&&&& \uparrow &&&&&&& \downarrow \\
    && \text{\textcolor{blue}{~$a_4$}} &&&&&&&&&& \text{\textcolor{blue}{~$a_3$}} &&&& \text{\textcolor{blue}{~$a_2$}} &&&&&& \text{\textcolor{blue}{~$a_1$}} &&&&&& \text{\textcolor{blue}{~$a_0$}} &&&&& &&\text{\textcolor{blue}{$P_4(x)\,$}}
    \end{myqcircuit*}$}
    } \\
  \caption{\eh\ circuits for a degree-4 polynomial $P_4(x) = \frac{1}{5}(a_0+a_1x+a_2x^2+a_3x^3+a_4x^4)$: (a) \emph{Reversible} 12-qubits \eh\ circuit for computing $P_4(x)$, where the  4 copies of the input $x$ are stored in  qubits $d_i$, and the 5 polynomial coefficients  are encoded in qubits $a_i$. The 3 ancilla qubits $\text{anc}_i$ are needed for concatenation of the 4 sum operators. Note that only qubit $a_0$ is measured. Unitaries labeled as $\Pi$ and $\Sigma_w$ represent the multiplication and addition circuits shown in \cref{fig:eh-prod,fig:eh-sum}, respectively. (b) \emph{Non-reversible} 5-qubits \eh\ circuit for the same degree-4 polynomial using only 1 ancilla qubit and  many resets allowing for qubit
recycling.}
    \label{fig:eh-poly_4}
\end{figure*}

\section{Polynomial construction}

Based on the four arithmetic quantum operations of \eh, we present a constructive method for computing a degree-$d$ polynomial, $P_d(x)$, on a quantum state. The polynomial is given by:
\begin{equation}
P_d(x) = \frac{1}{d+1} \left(a_0 + a_1x + a_2x^2 + \cdots + a_dx^d\right),
\label{eq:poly}
\end{equation}
where the input $x$, the polynomial coefficients $a_i$, and the output are all real values constrained to the range $[-1,1]$.

The $d$ copies of the real-valued input $x$ are encoded in an EVEN quantum product state on $d$ qubits.
To compute $P_d(x)$, we construct a quantum algorithm that sequentially applies addition and multiplication operations, similar to how a polynomial is evaluated on a classical calculator. 
There are multiple valid approaches to achieve this construction. Our reference implementation proceeds as follows: (i) compute all powers of the input $x$, (ii) multiply each power $x^k$ by its corresponding coefficient $a_k$, and (iii) sum the terms sequentially.
Since we employ the \emph{dyadic} operation $S_w(a,b)$, which combines only two inputs at a time, a sequence of $d$ such operations is needed to obtain $P_d(x)$.

A complete reversible quantum circuit for computing a degree $d = 4$ polynomial on a real number $x$ is illustrated in~\cref{fig:eh-poly_4-rev}. Due to the \emph{no-cloning} theorem, which prohibits reusing quantum states in multiple operations, four separate copies of the input are required.
Since the multiplication operator does not alter the upper qubit, the information stored there can be reused twice, as shown in \cref{eq:x0q0} of \cref{secSM:EHands}. 
The 5 user-defined polynomial coefficients stored in qubits $a_i$ are likewise encoded in an EVEN quantum product state. To ensure proper cancellation of unwanted quantum state components, 3 parity flip operations must be applied between successive sum operators. This requires 3 ancilla qubits prepared in superposition, which act as a fair-coin random generators through the applications of CZ gates.

\cref{fig:eh-poly_4-irrev} shows the non-reversible version of the same $P_4(x)$. The quantum resources required for the reversible and non-reversible implementations of a polynomial $P_d(x)$ are presented in~\cref{tab:cost}. \Cref{secSM:shallow} showcases yet another implementation of $P_d(x)$ requiring  only $\mathcal{O}(\log d)$ depth of CNOT gates.

\begin{table}[h!]%
\caption{\label{tab:cost} Resources and connectivity needed for reversible and non-reversible implementations of a polynomial $P_d(x)$. Both implementations require one final qubit measurement. 
}
\begin{ruledtabular}
\begin{tabular}{lccccc}
\textrm{type} &
\textrm{qubits} &
\textrm{ancilla} &
\textrm{resets} &
\textrm{CNOTs} &
\textrm{connectivity}\\
\colrule
reversible   &  3$d$ & $d-1$ & 0 & $5d-2$  & limited\\
non-reversible  &  $d+1$ & 0 &  $2d-1$ & $5d-2$ & linear\\
\end{tabular}
\end{ruledtabular}
\end{table}

\textit{A priori}, it is not obvious which of the two versions—reversible or non-reversible—will provide higher-fidelity execution on NISQ-era hardware. There is a trade-off between the noise introduced by the resets versus using a larger number of qubits, some of which may experience long idle times. Also a  more demanding connectivity may require additional SWAP operations.
Only the reversible \eh{} can be used as an \textit{oracle} for Grover-type search algorithms. 
However, one can simulate much higher-degree polynomials using the non-reversible version, since classical simulation memory requirements scale exponentially with qubit count.

\section{Quantum experiments}

To validate the \ehp\ on real quantum hardware, we selected  the three functions listed in~\cref{tab:ibm} and executed the corresponding \eh\ circuits on IBM's \emph{Aachen, Kingston}, and \emph{Marrakesh}.

Each target function was approximated by a polynomial of the desired degree. During classical pre-processing, the function was tabulated, fitted with a degree-$d$ polynomial, and the resulting coefficients were rescaled so that their maximum absolute value did not exceed 1. The resulting polynomials were then encoded using the \ehp.

\begin{table}[h!]
\caption{\label{tab:ibm}Polynomial approximations of functions computed on real IBM  quantum hardware. 
}
\begin{ruledtabular}
\begin{tabular}{lccccc}
\textrm{target} &
\textrm{\cref{fig:ibm}} &
\textrm{poly} &
\textrm{num} &
\textrm{num}&
\textrm{used} \\ 
\textrm{function} &
\textrm{panel} &
\textrm{degree} &
\textrm{CNOTs\footnote{circuit transpiled for heavy-hex qubit connectivity}} &
\textrm{qubits}&
\textrm{shots\footnote{number of shots per $x$-data point}} \\
\colrule
$\mathrm{ReLU}(x/2)$ & a & 3 & 28 & 9 & 2e4\\
$\arctan(5x)$        & b & 3 & 28 & 9 & 9e4\\
$x^2/3$              & c & 2 & 14 & 6 & 9e4\\
\end{tabular}
\end{ruledtabular}
\end{table}

\begin{figure*}[htbp]
\centering
\includegraphics[width=.91\linewidth]{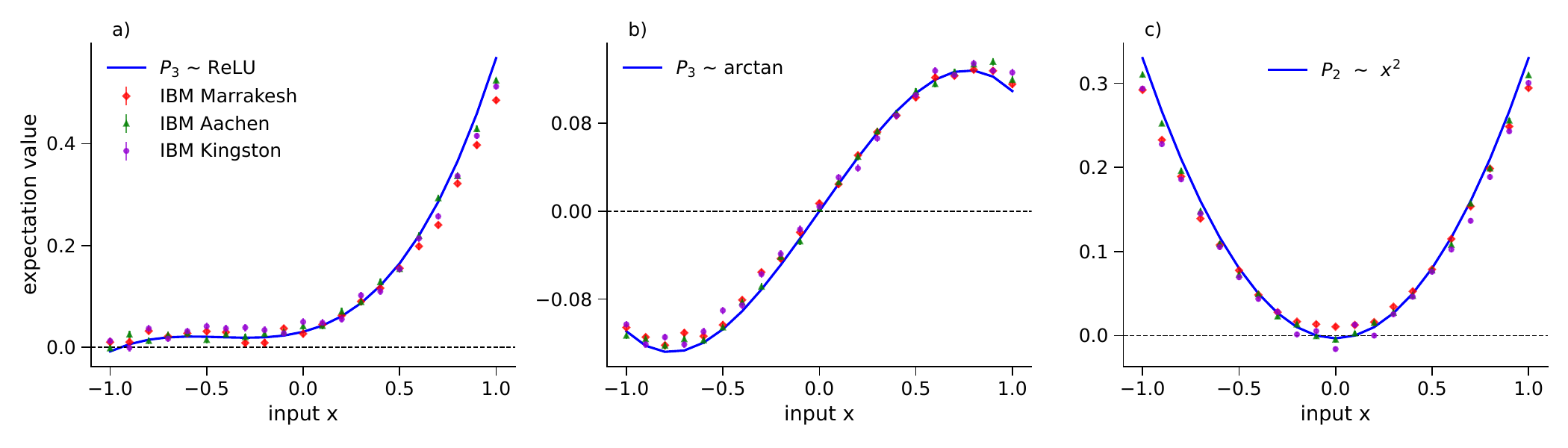}
   \caption{Hardware results from IBM's \emph{Aachen, Kingston}, and \emph{Marrakesh} for polynomial approximations of the functions listed in \cref{tab:ibm}, where we enabled Pauli twirling to mitigate hardware errors.}
\label{fig:ibm}
\end{figure*}

The results, shown in~\cref{fig:ibm}, demonstrate good agreement with the target functions. 
To mitigate hardware errors, we employed randomized compilation~\cite{PhysRevA.94.052325} and optimized qubit mapping~\cite{PRXQuantum.4.010327}, while omitting dynamical decoupling and zero noise extrapolation. \Cref{secSM:hardware} contains additional experimental results with each method toggled on and off, and provides a detailed justification for our decisions.
The additional simulations for different polynomials are included in \cref{secSM:simulations}.
The remaining discrepancies between the results from different QPUs reflect the individual biases introduced by each NISQ device. 

\section{Discussion}

The \ehp\ offers a versatile framework for implementing polynomial computations on QPUs, especially on NISQ devices. By leveraging standard quantum gates, our protocol implements efficient multivariable polynomial approximations for data stored in the EVEN-encoded initial state. 
In comparison to the quantum singular value transformation (QSVT) method, which implements polynomial transformations of the singular values of a block-encoded matrix, \eh\ computes the polynomial directly on the input data and therefore avoids the need to interleave the signal operator containing the input data with unitary operators. The closest analogue to \eh\ is the QSP protocol, which, to compute an arbitrary-parity polynomial of degree $d$, requires 3 qubits and roughly $12d$ two-qubit entangling gates (see \cref{secSM:QSP}). By comparison, \eh\ uses $3d$ qubits and only $5d-2$ two-qubit gates.
By directly operating on real-valued data encoded in quantum states, \eh\ is well-suited for applications with naturally available real-number inputs, such as sensor data or machine learning features. Experimental results on current quantum hardware demonstrate solid accuracy of the \ehp\ when standard error mitigation techniques are applied.
As hardware performance improves, we anticipate that even deeper circuits for higher-degree polynomials will yield reliable results.

A key advantage of the \ehp{} is its capability to perform \emph{vectorized} polynomial operations. Due to space constraints, this aspect is not explored in detail here.
However, by incorporating advanced data encoding schemes such as QPIXL~\cite{qpixl} and \qcrank~\cite{qbart}, a series of $n$ real-valued data points can be efficiently stored on a QPU using only $\lceil \log_2 n \rceil$ additional address qubits. 
This integration enables \emph{parallel computation} of the desired polynomial across entire sequences of real numbers, making \eh\ particularly useful for non-stationary signal analysis and noise/interference reduction directly on quantum hardware. 
A carefully chosen polynomial can adapt to varying signal characteristics, enhancing accuracy in dynamic environments. 

The ability to perform batch processing using vectorized quantum operations makes the \ehp\ a valuable tool for tasks such as computing batch-average mean square error (MSE) on a QPU in ML-regression problems. 
Notably, the MSE, when measured as the expectation value of a single qubit, serves as a local cost function, helping to mitigate the problem of barren plateaus. The details of the MSE implementation 
are discussed in \cref{secSM:MSE}.

In quantum image processing (QImP)~\cite{VenegasAndraca2003QIMP,10.1145/3663577,Wei2023} polynomial-based filters play a crucial role in edge detection and noise reduction.
By integrating the \ehp{} with efficient image encoding strategies, such as \qcrank{}, these image processing tasks can be performed solely on the quantum device. 
This integration could drive significant advancements in quantum-enhanced image analysis pipelines.

Note that the performance of the \eh\ algorithm is currently evaluated using the EVEN encoding. However, future quantum sensors are expected to interface directly with QPUs while preserving quantum information~\cite{Awschalometal2021,Sekiguchi2021,Chen2023}, thereby eliminating the need to encode classical input data through the EVEN scheme.
Such direct integration would significantly enhance the practical applicability of the \eh{} algorithm.

Despite its flexibility and efficiency, the \ehp{} faces scalability challenges when measuring expectation values for higher-degree polynomials. Due to signal attenuation of \( \mathcal{O}(1/d) \), achieving accurate results requires increasing the number of measurements, which scale as \( \mathcal{O}(d^2) \) to maintain the precision. While increasing shot counts can partially address this issue, exploring amplitude amplification techniques may provide more efficient solutions.

In conclusion, \eh\ represents a significant step forward in leveraging quantum hardware for real-world applications that rely on polynomial computations.
Its potential spans multiple domains, including QML, signal processing, and image analysis, where efficient polynomial transformations are essential. 
As quantum hardware continues to advance, the scalability and adaptability of the \ehp{} make it a promising candidate for further development and integration into more complex quantum computing frameworks.

For the sake of reproducibility and transparency, all circuits and experimental results used in this paper are publicly available in the repository \url{https://github.com/QuantumComputingLab/qpixlpp/} in the \texttt{examples} folder.

\begin{acknowledgments}
The authors are grateful to Daan Camps for helpful discussions.
This research was supported by the U.S. Department of Energy (DOE) under Contract No.~DE-AC02-05CH11231, through the Office of Science, Office of Advanced Scientific Computing Research (ASCR) Exploratory Research for Extreme-Scale Science and Accelerated Research in Quantum Computing.
This research used resources of two DOE user facilities: the National Energy Research Scientific Computing Center (NERSC) located at Lawrence Berkeley National Laboratory, operated under Contract No.~DE-AC02-05CH11231, and the Oak Ridge Leadership Computing Facility, operated under Contract No.~DE-AC05-00OR22725.
\end{acknowledgments}

\bibliography{bibliography}

\begin{thebibliography}{34}%
\makeatletter
\providecommand \@ifxundefined [1]{%
 \@ifx{#1\undefined}
}%
\providecommand \@ifnum [1]{%
 \ifnum #1\expandafter \@firstoftwo
 \else \expandafter \@secondoftwo
 \fi
}%
\providecommand \@ifx [1]{%
 \ifx #1\expandafter \@firstoftwo
 \else \expandafter \@secondoftwo
 \fi
}%
\providecommand \natexlab [1]{#1}%
\providecommand \enquote  [1]{``#1''}%
\providecommand \bibnamefont  [1]{#1}%
\providecommand \bibfnamefont [1]{#1}%
\providecommand \citenamefont [1]{#1}%
\providecommand \href@noop [0]{\@secondoftwo}%
\providecommand \href [0]{\begingroup \@sanitize@url \@href}%
\providecommand \@href[1]{\@@startlink{#1}\@@href}%
\providecommand \@@href[1]{\endgroup#1\@@endlink}%
\providecommand \@sanitize@url [0]{\catcode `\\12\catcode `\$12\catcode
  `\&12\catcode `\#12\catcode `\^12\catcode `\_12\catcode `\%12\relax}%
\providecommand \@@startlink[1]{}%
\providecommand \@@endlink[0]{}%
\providecommand \url  [0]{\begingroup\@sanitize@url \@url }%
\providecommand \@url [1]{\endgroup\@href {#1}{\urlprefix }}%
\providecommand \urlprefix  [0]{URL }%
\providecommand \Eprint [0]{\href }%
\providecommand \doibase [0]{https://doi.org/}%
\providecommand \selectlanguage [0]{\@gobble}%
\providecommand \bibinfo  [0]{\@secondoftwo}%
\providecommand \bibfield  [0]{\@secondoftwo}%
\providecommand \translation [1]{[#1]}%
\providecommand \BibitemOpen [0]{}%
\providecommand \bibitemStop [0]{}%
\providecommand \bibitemNoStop [0]{.\EOS\space}%
\providecommand \EOS [0]{\spacefactor3000\relax}%
\providecommand \BibitemShut  [1]{\csname bibitem#1\endcsname}%
\let\auto@bib@innerbib\@empty
\bibitem [{\citenamefont {Shor}(1994)}]{shor1994}%
  \BibitemOpen
  \bibfield  {author} {\bibinfo {author} {\bibfnamefont {P.~W.}\ \bibnamefont
  {Shor}},\ }\bibfield  {title} {\bibinfo {title} {{Algorithms for quantum
  computation: discrete logarithms and factoring}},\ }in\ \href
  {https://doi.org/10.1109/SFCS.1994.365700} {\emph {\bibinfo {booktitle}
  {Proceedings 35th Annual Symposium on Foundations of Computer Science}}}\
  (\bibinfo {year} {1994})\ pp.\ \bibinfo {pages} {124--134}\BibitemShut
  {NoStop}%
\bibitem [{\citenamefont {Grover}(1996)}]{grov1996}%
  \BibitemOpen
  \bibfield  {author} {\bibinfo {author} {\bibfnamefont {L.~K.}\ \bibnamefont
  {Grover}},\ }\bibfield  {title} {\bibinfo {title} {{A fast quantum mechanical
  algorithm for database search}},\ }in\ \href
  {https://doi.org/10.1145/237814.237866} {\emph {\bibinfo {booktitle}
  {Proceedings of the 28th Annual ACM Symposium on Theory of Computing}}}\
  (\bibinfo {year} {1996})\ pp.\ \bibinfo {pages} {212--219}\BibitemShut
  {NoStop}%
\bibitem [{\citenamefont {Harrow}\ \emph {et~al.}(2009)\citenamefont {Harrow},
  \citenamefont {Hassidim},\ and\ \citenamefont {Lloyd}}]{haha2009}%
  \BibitemOpen
  \bibfield  {author} {\bibinfo {author} {\bibfnamefont {A.~W.}\ \bibnamefont
  {Harrow}}, \bibinfo {author} {\bibfnamefont {A.}~\bibnamefont {Hassidim}},\
  and\ \bibinfo {author} {\bibfnamefont {S.}~\bibnamefont {Lloyd}},\ }\bibfield
   {title} {\bibinfo {title} {{Quantum algorithm for linear systems of
  equations}},\ }\href {https://doi.org/10.1103/PhysRevLett.103.150502}
  {\bibfield  {journal} {\bibinfo  {journal} {Phys. Rev. Lett.}\ }\textbf
  {\bibinfo {volume} {103}},\ \bibinfo {pages} {150502} (\bibinfo {year}
  {2009})}\BibitemShut {NoStop}%
\bibitem [{\citenamefont {Low}\ and\ \citenamefont
  {Chuang}(2017{\natexlab{a}})}]{low2017optimal}%
  \BibitemOpen
  \bibfield  {author} {\bibinfo {author} {\bibfnamefont {G.~H.}\ \bibnamefont
  {Low}}\ and\ \bibinfo {author} {\bibfnamefont {I.~L.}\ \bibnamefont
  {Chuang}},\ }\bibfield  {title} {\bibinfo {title} {{Optimal Hamiltonian
  simulation by quantum signal processing}},\ }\href
  {https://doi.org/10.1103/PhysRevLett.118.010501} {\bibfield  {journal}
  {\bibinfo  {journal} {Phys. Rev. Lett.}\ }\textbf {\bibinfo {volume} {118}},\
  \bibinfo {pages} {010501} (\bibinfo {year} {2017}{\natexlab{a}})}\BibitemShut
  {NoStop}%
\bibitem [{\citenamefont {Vedral}\ \emph {et~al.}(1996)\citenamefont {Vedral},
  \citenamefont {Barenco},\ and\ \citenamefont {Ekert}}]{veba1996}%
  \BibitemOpen
  \bibfield  {author} {\bibinfo {author} {\bibfnamefont {V.}~\bibnamefont
  {Vedral}}, \bibinfo {author} {\bibfnamefont {A.}~\bibnamefont {Barenco}},\
  and\ \bibinfo {author} {\bibfnamefont {A.}~\bibnamefont {Ekert}},\ }\bibfield
   {title} {\bibinfo {title} {{Quantum networks for elementary arithmetic
  operations}},\ }\href {https://doi.org/10.1103/PhysRevA.54.147} {\bibfield
  {journal} {\bibinfo  {journal} {Phys. Rev. A}\ }\textbf {\bibinfo {volume}
  {54}},\ \bibinfo {pages} {147} (\bibinfo {year} {1996})}\BibitemShut
  {NoStop}%
\bibitem [{\citenamefont {Takahashi}\ and\ \citenamefont
  {Kunihiro}(2005)}]{taku2005}%
  \BibitemOpen
  \bibfield  {author} {\bibinfo {author} {\bibfnamefont {Y.}~\bibnamefont
  {Takahashi}}\ and\ \bibinfo {author} {\bibfnamefont {N.}~\bibnamefont
  {Kunihiro}},\ }\bibfield  {title} {\bibinfo {title} {{A linear-size quantum
  circuit for addition with no ancillary qubits}},\ }\href
  {https://doi.org/10.26421/QIC5.6-2} {\bibfield  {journal} {\bibinfo
  {journal} {Quantum Inf. Comput.}\ }\textbf {\bibinfo {volume} {5}},\ \bibinfo
  {pages} {440} (\bibinfo {year} {2005})}\BibitemShut {NoStop}%
\bibitem [{\citenamefont {Takahashi}\ and\ \citenamefont
  {Kunihiro}(2008)}]{taku2008}%
  \BibitemOpen
  \bibfield  {author} {\bibinfo {author} {\bibfnamefont {Y.}~\bibnamefont
  {Takahashi}}\ and\ \bibinfo {author} {\bibfnamefont {N.}~\bibnamefont
  {Kunihiro}},\ }\bibfield  {title} {\bibinfo {title} {{A fast quantum circuit
  for addition with few qubits}},\ }\href {https://doi.org/10.26421/QIC8.6-7-5}
  {\bibfield  {journal} {\bibinfo  {journal} {Quantum Inf. Comput.}\ }\textbf
  {\bibinfo {volume} {8}},\ \bibinfo {pages} {636} (\bibinfo {year}
  {2008})}\BibitemShut {NoStop}%
\bibitem [{\citenamefont {Takahashi}(2009)}]{taka2009}%
  \BibitemOpen
  \bibfield  {author} {\bibinfo {author} {\bibfnamefont {Y.}~\bibnamefont
  {Takahashi}},\ }\bibfield  {title} {\bibinfo {title} {{Quantum arithmetic
  circuits: A survey}},\ }\href {https://doi.org/10.1587/transfun.E92.A.1276}
  {\bibfield  {journal} {\bibinfo  {journal} {IEICE Trans. Fundam. Electron.
  Commun. Comput. Sci.}\ }\textbf {\bibinfo {volume} {E92.A}},\ \bibinfo
  {pages} {1276} (\bibinfo {year} {2009})}\BibitemShut {NoStop}%
\bibitem [{\citenamefont {Wang}\ \emph {et~al.}(2025)\citenamefont {Wang},
  \citenamefont {Li}, \citenamefont {Lee}, \citenamefont {Deb}, \citenamefont
  {Lim},\ and\ \citenamefont {Chattopadhyay}}]{wali2025}%
  \BibitemOpen
  \bibfield  {author} {\bibinfo {author} {\bibfnamefont {S.}~\bibnamefont
  {Wang}}, \bibinfo {author} {\bibfnamefont {X.}~\bibnamefont {Li}}, \bibinfo
  {author} {\bibfnamefont {W.~J.~B.}\ \bibnamefont {Lee}}, \bibinfo {author}
  {\bibfnamefont {S.}~\bibnamefont {Deb}}, \bibinfo {author} {\bibfnamefont
  {E.}~\bibnamefont {Lim}},\ and\ \bibinfo {author} {\bibfnamefont
  {A.}~\bibnamefont {Chattopadhyay}},\ }\bibfield  {title} {\bibinfo {title}
  {{A comprehensive study of quantum arithmetic circuits}},\ }\href
  {https://doi.org/10.1098/rsta.2023.0392} {\bibfield  {journal} {\bibinfo
  {journal} {Philos. Trans. R. Soc. A}\ }\textbf {\bibinfo {volume} {383}},\
  \bibinfo {pages} {20230392} (\bibinfo {year} {2025})}\BibitemShut {NoStop}%
\bibitem [{\citenamefont {Low}\ and\ \citenamefont
  {Chuang}(2017{\natexlab{b}})}]{loch2017}%
  \BibitemOpen
  \bibfield  {author} {\bibinfo {author} {\bibfnamefont {G.~H.}\ \bibnamefont
  {Low}}\ and\ \bibinfo {author} {\bibfnamefont {I.~L.}\ \bibnamefont
  {Chuang}},\ }\bibfield  {title} {\bibinfo {title} {{Optimal Hamiltonian
  simulation by quantum signal processing}},\ }\href
  {https://doi.org/10.1103/PhysRevLett.118.010501} {\bibfield  {journal}
  {\bibinfo  {journal} {Phys. Rev. Lett.}\ }\textbf {\bibinfo {volume} {118}},\
  \bibinfo {pages} {010501} (\bibinfo {year} {2017}{\natexlab{b}})}\BibitemShut
  {NoStop}%
\bibitem [{\citenamefont {Dong}\ \emph {et~al.}(2024)\citenamefont {Dong},
  \citenamefont {Lin}, \citenamefont {Ni},\ and\ \citenamefont
  {Wang}}]{doli2024}%
  \BibitemOpen
  \bibfield  {author} {\bibinfo {author} {\bibfnamefont {Y.}~\bibnamefont
  {Dong}}, \bibinfo {author} {\bibfnamefont {L.}~\bibnamefont {Lin}}, \bibinfo
  {author} {\bibfnamefont {H.}~\bibnamefont {Ni}},\ and\ \bibinfo {author}
  {\bibfnamefont {J.}~\bibnamefont {Wang}},\ }\bibfield  {title} {\bibinfo
  {title} {{Infinite quantum signal processing}},\ }\href
  {https://doi.org/10.22331/q-2024-12-10-1558} {\bibfield  {journal} {\bibinfo
  {journal} {Quantum}\ }\textbf {\bibinfo {volume} {8}},\ \bibinfo {pages}
  {1558} (\bibinfo {year} {2024})}\BibitemShut {NoStop}%
\bibitem [{\citenamefont {Gily{\'{e}}n}\ \emph {et~al.}(2019)\citenamefont
  {Gily{\'{e}}n}, \citenamefont {Su}, \citenamefont {Low},\ and\ \citenamefont
  {Wiebe}}]{gisu2019}%
  \BibitemOpen
  \bibfield  {author} {\bibinfo {author} {\bibfnamefont {A.}~\bibnamefont
  {Gily{\'{e}}n}}, \bibinfo {author} {\bibfnamefont {Y.}~\bibnamefont {Su}},
  \bibinfo {author} {\bibfnamefont {G.~H.}\ \bibnamefont {Low}},\ and\ \bibinfo
  {author} {\bibfnamefont {N.}~\bibnamefont {Wiebe}},\ }\bibfield  {title}
  {\bibinfo {title} {{Quantum singular value transformation and beyond:
  exponential improvements for quantum matrix arithmetics}},\ }in\ \href
  {https://doi.org/10.1145/3313276.3316366} {\emph {\bibinfo {booktitle}
  {Proceedings of the 51st Annual ACM SIGACT Symposium on Theory of
  Computing}}}\ (\bibinfo {year} {2019})\ pp.\ \bibinfo {pages}
  {193--204}\BibitemShut {NoStop}%
\bibitem [{\citenamefont {Dong}\ \emph {et~al.}(2022)\citenamefont {Dong},
  \citenamefont {Lin},\ and\ \citenamefont {Tong}}]{doli2022}%
  \BibitemOpen
  \bibfield  {author} {\bibinfo {author} {\bibfnamefont {Y.}~\bibnamefont
  {Dong}}, \bibinfo {author} {\bibfnamefont {L.}~\bibnamefont {Lin}},\ and\
  \bibinfo {author} {\bibfnamefont {Y.}~\bibnamefont {Tong}},\ }\bibfield
  {title} {\bibinfo {title} {{Ground-state preparation and energy estimation on
  early fault-tolerant quantum computers via quantum eigenvalue transformation
  of unitary matrices}},\ }\href {https://doi.org/10.1103/PRXQuantum.3.040305}
  {\bibfield  {journal} {\bibinfo  {journal} {PRX Quantum}\ }\textbf {\bibinfo
  {volume} {3}},\ \bibinfo {pages} {40305} (\bibinfo {year}
  {2022})}\BibitemShut {NoStop}%
\bibitem [{\citenamefont {Kessler}\ \emph {et~al.}(2023)\citenamefont
  {Kessler}, \citenamefont {Alonso},\ and\ \citenamefont
  {S\'anchez}}]{grover-shots}%
  \BibitemOpen
  \bibfield  {author} {\bibinfo {author} {\bibfnamefont {M.}~\bibnamefont
  {Kessler}}, \bibinfo {author} {\bibfnamefont {D.}~\bibnamefont {Alonso}},\
  and\ \bibinfo {author} {\bibfnamefont {P.}~\bibnamefont {S\'anchez}},\
  }\bibfield  {title} {\bibinfo {title} {Determination of the number of shots
  for grover's search algorithm},\ }\href
  {https://doi.org/10.1140/epjqt/s40507-023-00204-y} {\bibfield  {journal}
  {\bibinfo  {journal} {EPJ Quantum Technol.}\ }\textbf {\bibinfo {volume}
  {10}},\ \bibinfo {pages} {47} (\bibinfo {year} {2023})}\BibitemShut {NoStop}%
\bibitem [{\citenamefont {Liu}\ and\ \citenamefont
  {at~al}(2023)}]{phase-estimation-shots}%
  \BibitemOpen
  \bibfield  {author} {\bibinfo {author} {\bibfnamefont {L.-Z.}\ \bibnamefont
  {Liu}}\ and\ \bibinfo {author} {\bibnamefont {at~al}},\ }\bibfield  {title}
  {\bibinfo {title} {Full-period quantum phase estimation},\ }\href
  {https://doi.org/10.1103/PhysRevLett.130.120802} {\bibfield  {journal}
  {\bibinfo  {journal} {Phys. Rev. Lett.}\ }\textbf {\bibinfo {volume} {130}},\
  \bibinfo {pages} {120802} (\bibinfo {year} {2023})}\BibitemShut {NoStop}%
\bibitem [{\citenamefont {He}\ \emph {et~al.}(2025)\citenamefont {He},
  \citenamefont {Amaro}, \citenamefont {Shaydulin},\ and\ \citenamefont
  {Pistoia}}]{qaoa-shots}%
  \BibitemOpen
  \bibfield  {author} {\bibinfo {author} {\bibfnamefont {Z.}~\bibnamefont
  {He}}, \bibinfo {author} {\bibfnamefont {D.}~\bibnamefont {Amaro}}, \bibinfo
  {author} {\bibfnamefont {R.}~\bibnamefont {Shaydulin}},\ and\ \bibinfo
  {author} {\bibfnamefont {M.}~\bibnamefont {Pistoia}},\ }\bibfield  {title}
  {\bibinfo {title} {Performance of quantum approximate optimization with
  quantum error detection},\ }\href
  {https://doi.org/10.1038/s42005-025-02136-8} {\bibfield  {journal} {\bibinfo
  {journal} {Commun. Phys.}\ }\textbf {\bibinfo {volume} {8}},\ \bibinfo
  {pages} {217} (\bibinfo {year} {2025})}\BibitemShut {NoStop}%
\bibitem [{\citenamefont {Herbert}(2022)}]{qmc-shots}%
  \BibitemOpen
  \bibfield  {author} {\bibinfo {author} {\bibfnamefont {S.}~\bibnamefont
  {Herbert}},\ }\bibfield  {title} {\bibinfo {title} {Quantum monte carlo
  integration: The full advantage in minimal circuit depth},\ }\href
  {https://doi.org/10.22331/q-2022-09-29-823} {\bibfield  {journal} {\bibinfo
  {journal} {Quantum}\ }\textbf {\bibinfo {volume} {6}},\ \bibinfo {pages}
  {823} (\bibinfo {year} {2022})}\BibitemShut {NoStop}%
\bibitem [{\citenamefont {Abrahao}\ \emph {et~al.}(2024)\citenamefont
  {Abrahao}, \citenamefont {Mansouri},\ and\ \citenamefont
  {Lund}}]{boson-shots}%
  \BibitemOpen
  \bibfield  {author} {\bibinfo {author} {\bibfnamefont {R.~A.}\ \bibnamefont
  {Abrahao}}, \bibinfo {author} {\bibfnamefont {A.}~\bibnamefont {Mansouri}},\
  and\ \bibinfo {author} {\bibfnamefont {A.~P.}\ \bibnamefont {Lund}},\
  }\bibfield  {title} {\bibinfo {title} {Boson sampling with gaussian input
  states: Toward efficient scaling and certification},\ }\href
  {https://doi.org/10.1103/PhysRevA.110.052437} {\bibfield  {journal} {\bibinfo
   {journal} {Phys. Rev. A}\ }\textbf {\bibinfo {volume} {110}},\ \bibinfo
  {pages} {052437} (\bibinfo {year} {2024})}\BibitemShut {NoStop}%
\bibitem [{\citenamefont {Recio-Armengol}\ and\ \citenamefont
  {Bowles}(2025)}]{recioarmengol2025polynomial}%
  \BibitemOpen
  \bibfield  {author} {\bibinfo {author} {\bibfnamefont {E.}~\bibnamefont
  {Recio-Armengol}}\ and\ \bibinfo {author} {\bibfnamefont {J.}~\bibnamefont
  {Bowles}},\ }\bibfield  {title} {\bibinfo {title} {{IQPopt: Fast optimization
  of instantaneous quantum polynomial circuits in JAX}},\ }\bibfield  {journal}
  {\bibinfo  {journal} {arXiv:}\ }\href
  {https://doi.org/10.48550/arXiv.2501.04776} {10.48550/arXiv.2501.04776}
  (\bibinfo {year} {2025})\BibitemShut {NoStop}%
\bibitem [{\citenamefont {Rath}\ and\ \citenamefont {Date}(2024)}]{rada2024}%
  \BibitemOpen
  \bibfield  {author} {\bibinfo {author} {\bibfnamefont {M.}~\bibnamefont
  {Rath}}\ and\ \bibinfo {author} {\bibfnamefont {H.}~\bibnamefont {Date}},\
  }\bibfield  {title} {\bibinfo {title} {{Quantum data encoding: a comparative
  analysis of classical-to-quantum mapping techniques and their impact on
  machine learning accuracy}},\ }\href
  {https://doi.org/10.1140/epjqt/s40507-024-00285-3} {\bibfield  {journal}
  {\bibinfo  {journal} {EPJ Quantum Technol.}\ }\textbf {\bibinfo {volume}
  {11}},\ \bibinfo {pages} {72} (\bibinfo {year} {2024})}\BibitemShut {NoStop}%
\bibitem [{\citenamefont {Wallman}\ and\ \citenamefont
  {Emerson}(2016)}]{PhysRevA.94.052325}%
  \BibitemOpen
  \bibfield  {author} {\bibinfo {author} {\bibfnamefont {J.~J.}\ \bibnamefont
  {Wallman}}\ and\ \bibinfo {author} {\bibfnamefont {J.}~\bibnamefont
  {Emerson}},\ }\bibfield  {title} {\bibinfo {title} {Noise tailoring for
  scalable quantum computation via randomized compiling},\ }\href
  {https://doi.org/10.1103/PhysRevA.94.052325} {\bibfield  {journal} {\bibinfo
  {journal} {Phys. Rev. A}\ }\textbf {\bibinfo {volume} {94}},\ \bibinfo
  {pages} {052325} (\bibinfo {year} {2016})}\BibitemShut {NoStop}%
\bibitem [{\citenamefont {Nation}\ and\ \citenamefont
  {Treinish}(2023)}]{PRXQuantum.4.010327}%
  \BibitemOpen
  \bibfield  {author} {\bibinfo {author} {\bibfnamefont {P.~D.}\ \bibnamefont
  {Nation}}\ and\ \bibinfo {author} {\bibfnamefont {M.}~\bibnamefont
  {Treinish}},\ }\bibfield  {title} {\bibinfo {title} {Suppressing quantum
  circuit errors due to system variability},\ }\href
  {https://doi.org/10.1103/PRXQuantum.4.010327} {\bibfield  {journal} {\bibinfo
   {journal} {PRX Quantum}\ }\textbf {\bibinfo {volume} {4}},\ \bibinfo {pages}
  {010327} (\bibinfo {year} {2023})}\BibitemShut {NoStop}%
\bibitem [{\citenamefont {Amankwah}\ \emph {et~al.}(2022)\citenamefont
  {Amankwah}, \citenamefont {Camps}, \citenamefont {Bethel}, \citenamefont
  {{Van Beeumen}},\ and\ \citenamefont {Perciano}}]{qpixl}%
  \BibitemOpen
  \bibfield  {author} {\bibinfo {author} {\bibfnamefont {M.~G.}\ \bibnamefont
  {Amankwah}}, \bibinfo {author} {\bibfnamefont {D.}~\bibnamefont {Camps}},
  \bibinfo {author} {\bibfnamefont {E.~W.}\ \bibnamefont {Bethel}}, \bibinfo
  {author} {\bibfnamefont {R.}~\bibnamefont {{Van Beeumen}}},\ and\ \bibinfo
  {author} {\bibfnamefont {T.}~\bibnamefont {Perciano}},\ }\bibfield  {title}
  {\bibinfo {title} {{Quantum pixel representations and compression for
  N-dimensional images}},\ }\href {https://doi.org/10.1038/s41598-022-11024-y}
  {\bibfield  {journal} {\bibinfo  {journal} {Sci. Rep.}\ }\textbf {\bibinfo
  {volume} {12}},\ \bibinfo {pages} {7712} (\bibinfo {year}
  {2022})}\BibitemShut {NoStop}%
\bibitem [{\citenamefont {Balewski}\ \emph {et~al.}(2024)\citenamefont
  {Balewski}, \citenamefont {Amankwah}, \citenamefont {{Van Beeumen}},
  \citenamefont {Bethel}, \citenamefont {Perciano},\ and\ \citenamefont
  {Camps}}]{qbart}%
  \BibitemOpen
  \bibfield  {author} {\bibinfo {author} {\bibfnamefont {J.}~\bibnamefont
  {Balewski}}, \bibinfo {author} {\bibfnamefont {M.~G.}\ \bibnamefont
  {Amankwah}}, \bibinfo {author} {\bibfnamefont {R.}~\bibnamefont {{Van
  Beeumen}}}, \bibinfo {author} {\bibfnamefont {E.~W.}\ \bibnamefont {Bethel}},
  \bibinfo {author} {\bibfnamefont {T.}~\bibnamefont {Perciano}},\ and\
  \bibinfo {author} {\bibfnamefont {D.}~\bibnamefont {Camps}},\ }\bibfield
  {title} {\bibinfo {title} {{Quantum-parallel vectorized data encodings and
  computations on trapped-ion and transmon QPUs}},\ }\href
  {https://doi.org/10.1038/s41598-024-53720-x} {\bibfield  {journal} {\bibinfo
  {journal} {Sci. Rep.}\ }\textbf {\bibinfo {volume} {14}},\ \bibinfo {pages}
  {3435} (\bibinfo {year} {2024})}\BibitemShut {NoStop}%
\bibitem [{\citenamefont {Venegas-Andraca}\ and\ \citenamefont
  {Bose}(2003)}]{VenegasAndraca2003QIMP}%
  \BibitemOpen
  \bibfield  {author} {\bibinfo {author} {\bibfnamefont {S.~E.}\ \bibnamefont
  {Venegas-Andraca}}\ and\ \bibinfo {author} {\bibfnamefont {S.}~\bibnamefont
  {Bose}},\ }\bibfield  {title} {\bibinfo {title} {{Storing, processing, and
  retrieving an image using quantum mechanics}},\ }in\ \href
  {https://doi.org/10.1117/12.485960} {\emph {\bibinfo {booktitle} {Quantum
  Information and Computation}}},\ Vol.\ \bibinfo {volume} {5105}\ (\bibinfo
  {year} {2003})\ pp.\ \bibinfo {pages} {137--147}\BibitemShut {NoStop}%
\bibitem [{\citenamefont {Yan}\ and\ \citenamefont
  {Venegas-Andraca}(2025)}]{10.1145/3663577}%
  \BibitemOpen
  \bibfield  {author} {\bibinfo {author} {\bibfnamefont {F.}~\bibnamefont
  {Yan}}\ and\ \bibinfo {author} {\bibfnamefont {S.~E.}\ \bibnamefont
  {Venegas-Andraca}},\ }\bibfield  {title} {\bibinfo {title} {Lessons from
  twenty years of quantum image processing},\ }\href
  {https://doi.org/10.1145/3663577} {\bibfield  {journal} {\bibinfo  {journal}
  {ACM Trans. Quantum Comput.}\ }\textbf {\bibinfo {volume} {6}},\ \bibinfo
  {pages} {1} (\bibinfo {year} {2025})}\BibitemShut {NoStop}%
\bibitem [{\citenamefont {Wei}\ \emph {et~al.}(2023)\citenamefont {Wei},
  \citenamefont {Liu}, \citenamefont {Xu}, \citenamefont {Shi}, \citenamefont
  {Shan}, \citenamefont {Zhao},\ and\ \citenamefont {Gao}}]{Wei2023}%
  \BibitemOpen
  \bibfield  {author} {\bibinfo {author} {\bibfnamefont {L.}~\bibnamefont
  {Wei}}, \bibinfo {author} {\bibfnamefont {H.}~\bibnamefont {Liu}}, \bibinfo
  {author} {\bibfnamefont {J.}~\bibnamefont {Xu}}, \bibinfo {author}
  {\bibfnamefont {L.}~\bibnamefont {Shi}}, \bibinfo {author} {\bibfnamefont
  {Z.}~\bibnamefont {Shan}}, \bibinfo {author} {\bibfnamefont {B.}~\bibnamefont
  {Zhao}},\ and\ \bibinfo {author} {\bibfnamefont {Y.}~\bibnamefont {Gao}},\
  }\bibfield  {title} {\bibinfo {title} {Quantum machine learning in medical
  image analysis: A survey},\ }\href
  {https://doi.org/https://doi.org/10.1016/j.neucom.2023.01.049} {\bibfield
  {journal} {\bibinfo  {journal} {Neurocomputing}\ }\textbf {\bibinfo {volume}
  {525}},\ \bibinfo {pages} {42} (\bibinfo {year} {2023})}\BibitemShut
  {NoStop}%
\bibitem [{\citenamefont {Awschalom}\ and\ \citenamefont
  {at~al}(2021)}]{Awschalometal2021}%
  \BibitemOpen
  \bibfield  {author} {\bibinfo {author} {\bibfnamefont {D.}~\bibnamefont
  {Awschalom}}\ and\ \bibinfo {author} {\bibnamefont {at~al}},\ }\bibfield
  {title} {\bibinfo {title} {{Development of quantum interconnects (QuICs) for
  next-generation information technologies}},\ }\href
  {https://doi.org/10.1103/PRXQuantum.2.017002} {\bibfield  {journal} {\bibinfo
   {journal} {PRX Quantum}\ }\textbf {\bibinfo {volume} {2}},\ \bibinfo {pages}
  {017002} (\bibinfo {year} {2021})}\BibitemShut {NoStop}%
\bibitem [{\citenamefont {Sekiguchi}\ \emph {et~al.}(2021)\citenamefont
  {Sekiguchi}, \citenamefont {Yasui}, \citenamefont {Tsurumoto} \emph
  {et~al.}}]{Sekiguchi2021}%
  \BibitemOpen
  \bibfield  {author} {\bibinfo {author} {\bibfnamefont {Y.}~\bibnamefont
  {Sekiguchi}}, \bibinfo {author} {\bibfnamefont {Y.}~\bibnamefont {Yasui}},
  \bibinfo {author} {\bibfnamefont {K.}~\bibnamefont {Tsurumoto}}, \emph
  {et~al.},\ }\bibfield  {title} {\bibinfo {title} {Geometric entanglement of a
  photon and spin qubits in diamond},\ }\href
  {https://doi.org/10.1038/s42005-021-00767-1} {\bibfield  {journal} {\bibinfo
  {journal} {Commun. Phys.}\ }\textbf {\bibinfo {volume} {4}},\ \bibinfo
  {pages} {264} (\bibinfo {year} {2021})}\BibitemShut {NoStop}%
\bibitem [{\citenamefont {Chen}(2023)}]{Chen2023}%
  \BibitemOpen
  \bibfield  {author} {\bibinfo {author} {\bibfnamefont {Z.}~\bibnamefont
  {Chen}},\ }\bibfield  {title} {\bibinfo {title} {Transferring
  four-dimensional atomic states at one step between separated cavities},\
  }\href {https://doi.org/10.1088/1742-6596/2595/1/012002} {\bibfield
  {journal} {\bibinfo  {journal} {J. Phys. Conf. Ser.}\ }\textbf {\bibinfo
  {volume} {2595}},\ \bibinfo {pages} {012002} (\bibinfo {year}
  {2023})}\BibitemShut {NoStop}%
\bibitem [{\citenamefont {Evert}\ \emph {et~al.}(2025)\citenamefont {Evert},
  \citenamefont {Izquierdo}, \citenamefont {Sud}, \citenamefont {Hu},
  \citenamefont {Grabbe}, \citenamefont {Rieffel}, \citenamefont {Reagor},\
  and\ \citenamefont {Wang}}]{8lxc-lvv1}%
  \BibitemOpen
  \bibfield  {author} {\bibinfo {author} {\bibfnamefont {B.}~\bibnamefont
  {Evert}}, \bibinfo {author} {\bibfnamefont {Z.~G.}\ \bibnamefont
  {Izquierdo}}, \bibinfo {author} {\bibfnamefont {J.}~\bibnamefont {Sud}},
  \bibinfo {author} {\bibfnamefont {H.-Y.}\ \bibnamefont {Hu}}, \bibinfo
  {author} {\bibfnamefont {S.}~\bibnamefont {Grabbe}}, \bibinfo {author}
  {\bibfnamefont {E.~G.}\ \bibnamefont {Rieffel}}, \bibinfo {author}
  {\bibfnamefont {M.~J.}\ \bibnamefont {Reagor}},\ and\ \bibinfo {author}
  {\bibfnamefont {Z.}~\bibnamefont {Wang}},\ }\bibfield  {title} {\bibinfo
  {title} {Syncopated dynamical decoupling to suppress crosstalk in quantum
  circuits},\ }\href {https://doi.org/10.1103/8lxc-lvv1} {\bibfield  {journal}
  {\bibinfo  {journal} {Phys. Rev. Appl.}\ ,\ } (\bibinfo {year}
  {2025})}\BibitemShut {NoStop}%
\bibitem [{\citenamefont {Martyn}\ \emph {et~al.}(2021)\citenamefont {Martyn},
  \citenamefont {Rossi}, \citenamefont {Tan},\ and\ \citenamefont
  {Chuang}}]{Martyn2021GrandUnification}%
  \BibitemOpen
  \bibfield  {author} {\bibinfo {author} {\bibfnamefont {J.~M.}\ \bibnamefont
  {Martyn}}, \bibinfo {author} {\bibfnamefont {Z.~M.}\ \bibnamefont {Rossi}},
  \bibinfo {author} {\bibfnamefont {A.~K.}\ \bibnamefont {Tan}},\ and\ \bibinfo
  {author} {\bibfnamefont {I.~L.}\ \bibnamefont {Chuang}},\ }\bibfield  {title}
  {\bibinfo {title} {Grand unification of quantum algorithms},\ }\href
  {https://doi.org/10.1103/PRXQuantum.2.040203} {\bibfield  {journal} {\bibinfo
   {journal} {PRX Quantum}\ }\textbf {\bibinfo {volume} {2}},\ \bibinfo {pages}
  {040203} (\bibinfo {year} {2021})}\BibitemShut {NoStop}%
\bibitem [{\citenamefont {Chakraborty}(2024)}]{Chakraborty2024LCU}%
  \BibitemOpen
  \bibfield  {author} {\bibinfo {author} {\bibfnamefont {S.}~\bibnamefont
  {Chakraborty}},\ }\bibfield  {title} {\bibinfo {title} {Implementing any
  linear combination of unitaries on intermediate-term quantum computers},\
  }\href {https://doi.org/10.22331/q-2024-10-10-1496} {\bibfield  {journal}
  {\bibinfo  {journal} {Quantum}\ }\textbf {\bibinfo {volume} {8}},\ \bibinfo
  {pages} {1496} (\bibinfo {year} {2024})}\BibitemShut {NoStop}%
\bibitem [{\citenamefont {Motlagh}\ and\ \citenamefont
  {Wiebe}(2024)}]{Motlagh2024GQSP}%
  \BibitemOpen
  \bibfield  {author} {\bibinfo {author} {\bibfnamefont {D.}~\bibnamefont
  {Motlagh}}\ and\ \bibinfo {author} {\bibfnamefont {N.}~\bibnamefont
  {Wiebe}},\ }\bibfield  {title} {\bibinfo {title} {Generalized quantum signal
  processing},\ }\href {https://doi.org/10.1103/PRXQuantum.5.020368} {\bibfield
   {journal} {\bibinfo  {journal} {PRX Quantum}\ }\textbf {\bibinfo {volume}
  {5}},\ \bibinfo {pages} {020368} (\bibinfo {year} {2024})}\BibitemShut
  {NoStop}%
\end{thebibliography}%


\onecolumngrid
\appendix

\newfloat{circuit}{hbtp}{loc}
\newcounter{subcircuit}[circuit]
\newcounter{subcircuit@save}[circuit]
\renewcommand\thesubcircuit{\alph{subcircuit}}

\section*{Appendixes}

In the Appendixes, we discuss in detail the derivation of the \ehp\ presented in the main text.
In \cref{secSM:EHands}, we describe the EVEN encoding and decoding schemes together with the four key \eh{} quantum circuits implementing \emph{multiplication}, \emph{addition}, \emph{negation}, and \emph{parity flip}.
We further generalize the weighted sum circuit to compute the linear combination of three real numbers and extend the \ehp\ to multivariable polynomials.
\Cref{secSM:MSE} discusses the computation of the mean squared error (MSE) as a specific example.
\Cref{secSM:simulations} presents numerical simulations that complement the experimental results reported in the main text, while \cref{secSM:hardware} details additional exploratory experiments on IBM hardware that motivates our choice of error mitigation techniques.
\Cref{secSM:shallow} introduces an alternative implementation of $P_d(x)$ with CNOT depth scaling as $\mathcal{O}(\log d)$.
Finally, \cref{secSM:QSP} outlines the Quantum Signal Processing (QSP) method, explaining how it is combined with the Hadamard test and Linear Combination of Unitaries (LCU) technique to enable the measurement of arbitrary parity polynomials $P_d(x)$ on a QPU.

\section{State vectors and observables for \ehptabs}
\label{secSM:EHands}

\subsection{EVEN encoding and decoding of 2 real numbers on 2 qubits}
\label{secSM:encode-decode}

\subparagraph{Encoding.}
Let us consider 2 real values \( \{x_0,x_1\} \) in the range \([-1, 1]\). The encoding process involves mapping each value \( x_i \) to an angle \( \theta_i \):
\begin{equation}
    \theta_i = \arccos(x_i), \qquad i = 0,1.
    \label{eq:app-theta}
\end{equation}
These angles parameterize rotation gates \( \Ry(\theta_i) \), which are then applied to a 2-qubit quantum state initialized in \( \ket{00} \):
\begin{equation}
    \ket{\phi} = \left[ \Ry(\theta_0) \otimes \Ry(\theta_1) \right] \ket{00}
    = \begin{bmatrix}
        \cos(\nicefrac{\theta_0}{2}) \cos(\nicefrac{\theta_1}{2}) \\
        \cos(\nicefrac{\theta_0}{2}) \sin(\nicefrac{\theta_1}{2}) \\
        \sin(\nicefrac{\theta_0}{2}) \cos(\nicefrac{\theta_1}{2}) \\
        \sin(\nicefrac{\theta_0}{2}) \sin(\nicefrac{\theta_1}{2}) \\
    \end{bmatrix}
    = \frac{1}{2} \begin{bmatrix}
        \sqrt{1 + x_0} \sqrt{1 + x_1} \\
        \sqrt{1 + x_0} \sqrt{1 - x_1} \\
        \sqrt{1 - x_0} \sqrt{1 + x_1} \\
        \sqrt{1 - x_0} \sqrt{1 - x_1} \\
    \end{bmatrix},
    \label{eq:x2theta}
\end{equation}
where we applied the last equality \cref{eq:app-theta} and relationships between trigonometric and inverse trigonometric functions.
The resulting state \( \ket{\phi} \), shown also as \cref{circ:eh-init}\ref{circ:eh-init-encode}, \emph{encodes} the classical information about  \( \{x_0,x_1\} \).

\begin{circuit}[hbtp]
\centering
\subfloat[Encoding \( \{x_0,x_1\} \)\label{circ:eh-init-encode}]{%
\begin{minipage}[b][5em][t]{0.3\textwidth}
\qquad\mbox{%
\Qcircuit @C=0.7em @R=1.2em {
& \lstick{\ket{0}_0} & \gate{\Ry(\theta_0)} &\qw \\
& \lstick{\ket{0}_1} & \gate{\Ry(\theta_1)} & \qw
} } \hspace{-1.5ex} \raisebox{-1.65em}{ \(\Bigg\} \ket\phi \)}
\end{minipage}
}%
\hfill%
%
\subfloat[Decoding \(x_0\)\label{circ:eh-init-decode0}]{%
\begin{minipage}[b][5em][t]{0.3\textwidth}
\vspace{0.25ex}
\hspace{-10em}\raisebox{-1.65em}{\( \ket\phi \Bigg\{ \)}
\mbox{%
\Qcircuit @C=0.7em @R=2.2em {
 &\qw & \measuretab{Z} & \rstick{\hat{y}_0 = \text{EV}(\sigma_z \otimes I) \simeq x_0} \\
 & \qw & \qw
} }
\end{minipage}
}%
\hfill%
%
\subfloat[Decoding \(x_1\)\label{circ:eh-init-decode1}]{%
\begin{minipage}[b][5em][t]{0.3\textwidth}
\vspace{1.65ex}
\hspace{-10em}\raisebox{-1.65em}{\( \ket\phi \Bigg\{ \)}
\mbox{%
\Qcircuit @C=0.7em @R=2.2em {
 &\qw & \qw \\
 & \qw &\measuretab{Z} & \rstick{\hat{y}_1 = \text{EV}(I \otimes \sigma_z) \simeq x_1}
} }
\end{minipage}
}
\caption{\justifying\small (a) Encoding of  2 real numbers \( \{x_0,x_1\} \) in a 2-qubit state \( \ket{\phi} \). (b)--(c) Decoding is performed by measuring the expectation value of the Pauli-$Z$ operator on the corresponding qubit.}
\label{circ:eh-init}
\end{circuit} 

\subparagraph{Decoding.}
In order to retrieve either of the value \( \{x_0,x_1\} \) encoded in the state $\ket\phi$, we define two complementary operators
\begin{align}
    O_{s} &:=  \sigma_z  \otimes I =  \begin{bmatrix}
        1 & 0 & 0 & 0 \\
        0 & 1 & 0 & 0 \\
        0 & 0 & -1 & 0 \\
        0 & 0 & 0 & -1 \\
    \end{bmatrix}, &
    O_{p} &:=  I \otimes \sigma_z  =  \begin{bmatrix}
        1 & 0 & 0 & 0 \\
        0 & -1 & 0 & 0 \\
        0 & 0 & 1 & 0 \\
        0 & 0 & 0 & -1 \\
    \end{bmatrix},
\end{align}
which represent measurements of qubit $q_0$ or qubit $q_1$ in the $Z$-basis, respectively.
We measure expectation values (EV) of the operators $O_s, O_p$, yielding back the estimators of the input values
\begin{align}
     \bra{\phi} O_s \ket{\phi} &= \hat x_0, &
     \bra{\phi} O_p \ket{\phi} &= \hat x_1.
\end{align}
The \emph{decoding} quantum circuits are depicted in \cref{circ:eh-init-decode0,circ:eh-init-decode1}, where 
\raisebox{0.2em}{\Qcircuit @C=0.4em {& \qw & \measuretab{Z}}} 
denotes measuring the EV of the corresponding qubit in the $Z$-basis.

\subsection{Product-with-memory circuit}

After establishing a method to encode and decode a pair of real values using a 2-qubit system, we introduce a unitary $\Uprod$ to compute their product.
When $\Uprod$ is applied to the state $\ket{\phi}$, as defined in \cref{eq:x2theta}, it produces a new state $\ket{\psi_{\Pi}}$.  
The expectation value (EV) of $O_p$ for this state corresponds to the estimator of the product $x_0 x_1$.  

Let
\(
    \Rz(\nicefrac{\pi}{2}) = e^{-i\frac{\pi}{4}} \begin{bmatrix}
        1 & 0 \\ 0 & i
    \end{bmatrix}.
\)
We define the \emph{product}-unitary $\Uprod$ as follows
\begin{equation}
    \Uprod := \text{CNOT}_{0,1}\cdot \left[ I \otimes  \Rz(\nicefrac{\pi}{2}) \right] = e^{-i\frac{\pi}{4}} \begin{bmatrix}
        1 & 0 & 0 & 0 \\
        0 & i & 0 & 0 \\
        0 & 0 & 0 & i \\
        0 & 0 & 1 & 0 \\
    \end{bmatrix}.
\end{equation}
Next, by applying $\Uprod$ to the encoded state $\ket\phi$ given in \cref{eq:x2theta}, we obtain the state
\begin{equation}
    \ket{\psi_\Pi} = \Uprod \ket\phi = \frac{e^{-i\frac{\pi}{4}}}{2} \left[\begin{array}{r}
           \sqrt{1 + x_0} \sqrt{1 + x_1} \\
        i\,\sqrt{1 + x_0} \sqrt{1 - x_1} \\
        i\,\sqrt{1 - x_0} \sqrt{1 - x_1} \\
           \sqrt{1 - x_0} \sqrt{1 + x_1} \\
    \end{array}\right],
\end{equation}
so that the EV of $O_p$ returns the product of the 2 inputs 
\begin{equation}
 \bra{\psi_\Pi}O_{p}\ket{\psi_\Pi} = x_0 x_1.
 \label{eq:PiOpPi}
\end{equation}
The corresponding quantum \cref{circ:eh-prod} performs $\hat{y}_1 =  x_0 x_1$. It includes the initial state preparation, where the rotation angles $\theta_i$ are computed as in~\cref{eq:app-theta}.

\begin{circuit}[hbtp]
\centering
\hspace{-10em}\mbox{%
\Qcircuit @C=0.7em @R=1.2em {
& & & & \quad\Uprod \\
& \lstick{\ket{0}_0} & \gate{\Ry(\theta_0)} & \qw & \qw & \ctrl{1} & \qw & \qw \\
& \lstick{\ket{0}_1} & \gate{\Ry(\theta_1)} & \qw & \gate{\Rz(\frac{\pi}{2})} & \targ & \qw & \measuretab{Z} & \rstick{\hat{y}_1 = \text{EV}(I \otimes \sigma_z) \simeq x_0 x_1}
{\gategroup{2}{5}{3}{6}{1.2em}{.}}
}}
\caption{\small Computing  product $x_0x_1$ as the expectation value of $\sigma_z$ operator measured on $q_1$. }
\label{circ:eh-prod}
\end{circuit}

Note that $\Uprod$ does not affect qubit $q_0$.  
As a result, this qubit retains the information about its input value $x_0$. Hence, the EV of $O_s$ after applying $\Uprod$ remains unchanged
\begin{equation}
 \bra{\psi_\Pi} O_s\ket{\psi_\Pi} = x_0.
 \label{eq:x0q0}
\end{equation}

For a general input vector of length $N$, the EVEN encoding requires $N$ qubits for data storage. 
The product-with-memory operation scales linearly with the number of input pairs, requiring $\mathcal{O}(N)$ CNOT gates with a circuit depth of $\mathcal{O}(1)$ for parallel multiplication operations, or $\mathcal{O}(N)$ for sequential operations depending on the specific implementation and connectivity constraints of the target quantum processor.

\subsection{Weighted sum circuit}

Our next goal is to compute linear combinations of a pair of real numbers.  
These numbers are again encoded in the 2-qubit state $\ket{\phi}$, as defined in \cref{eq:x2theta}.  
To do so, we introduce the unitary $\Usum$.  
When $\Usum$ is applied to the state $\ket{\phi}$,  it produces a new state $\ket{\psi_{\Sigma}}$.  
The expectation value (EV) of $O_s$ for this state corresponds 
 to the estimator of the weighted sum $w x_0 + (1-w) x_1$, where the relative weight $w \in [0,1]$.
\Cref{circ:eh-sum} encapsulates this entire process.

\begin{circuit}[hbtp] 
\centering
\mbox{%
\hspace{-15em}\Qcircuit @C=0.7em @R=1.2em {
& & & & & & \Usum(w) \\
& \lstick{\ket{0}_0} & \gate{\Ry(\theta_0)} & \qw & \qw & \ctrl{1} & \gate{\Ry(\frac{\alpha}{2})} & \targ & \gate{\Ry(-\frac{\alpha}{2})} & \qw & \measuretab{Z} & \rstick{\hat{y}_0 = \text{EV}(\sigma_z \otimes I) \simeq wx_0+(1-w)x_1} \\
& \lstick{\ket{0}_1} & \gate{\Ry(\theta_1)} & \qw & \gate{\Rz(\frac{\pi}{2})} & \targ & \qw & \ctrl{-1} & \qw & \qw & \qw
{\gategroup{2}{5}{3}{9}{1.2em}{.}}
}} 
\caption{\small Computing weighted sum $wx_0+(1-w)x_1$ as EV of $\sigma_z$ operator measured on $q_0$.}
\label{circ:eh-sum}
\end{circuit}

The unitary $\Usum$ contains a rotation parameter $\alpha$ which depends on $w$ as follows
\begin{equation}
    \alpha = \arccos(1-2w).
\end{equation}

To prove the above, we start by expressing the rotation matrices $\Ry(\nicefrac{\alpha}{2})$ and $\Ry(\nicefrac{-\alpha}{2})$, with $\alpha \in [0,\pi]$, in terms of the parameter $w$:
\begin{equation}
      \Ry(\pm\ \nicefrac{\alpha}{2}) = \frac{1}{\sqrt{2}}\left[\begin{array}{rr}
        \sqrt{1+\sqrt{1-w}} & \mp\sqrt{1-\sqrt{1-w}} \\\pm \sqrt{1-\sqrt{1-w}} & \sqrt{1+\sqrt{1-w}}
    \end{array}\right], 
\end{equation}
and define the parameterized \emph{sum}-unitary $\Usum(w)$ as follows
\begin{equation}
    \Usum(w) :=  \left[   \Ry(\nicefrac{-\alpha}{2})  \otimes I\right] \cdot \CNOT_{1,0}\cdot  \left[   \Ry(\nicefrac{\alpha}{2})  \otimes I\right] \cdot \Uprod = e^{-i\frac{\pi}{4}} \begin{bmatrix}
        1 & 0 & 0 & 0 \\
        0 & i\sqrt{w} & \sqrt{1-w}  & 0 \\
        0 & 0 & 0 & i \\
        0 & i \sqrt{1-w}  & -   \sqrt{w} & 0 \\
    \end{bmatrix}.
    \label{eq:Usum}
\end{equation}
Next, by applying $\Usum(w)$ to the encoded state $\ket\phi$ given in \cref{eq:x2theta}, we obtain the state
\begin{equation}
    \ket{\psi_\Sigma} = \Usum(w) \ket\phi = \frac{e^{-i\frac{\pi}{4}}}{2} \left[\begin{array}{r}
           \sqrt{1 + x_0} \sqrt{1 + x_1} \\
        i\,\sqrt{w}\sqrt{1 + x_0} \sqrt{1 - x_1} + \sqrt{1-w}\sqrt{1 - x_0} \sqrt{1 + x_1}\\
        i\,\sqrt{1 - x_0} \sqrt{1 - x_1} \\
           i\,\sqrt{1-w}\sqrt{1 + x_0} \sqrt{1 - x_1} -\sqrt{w}\sqrt{1 - x_0} \sqrt{1 + x_1} \\
    \end{array}\right],
\end{equation}
so that the EV of $O_s$ returns the weighted sum of the 2 inputs
\begin{equation}
    \bra{\psi_\Sigma}O_{s}\ket{\psi_\Sigma} = wx_0+(1-w)x_1.
    \label{eq:wsum}
\end{equation}
Note that the first 2 gates of $\Usum$ are equal to $\Uprod$ and the remaining gates of $\Usum$ do not change $q_1$ in \cref{circ:eh-sum}, hence,  qubit $q_1$ contains the $x_0 x_1$ product, i.e., 
\begin{equation}
    \bra{\psi_\Sigma}O_{p}\ket{\psi_\Sigma} = x_0 x_1.
\end{equation}
For the special choice of $ w=1/2 \rightarrow \alpha=\pi/2$ the result of \cref{eq:wsum} is a plain average: $ (x_0+x_1)/2$.

For completeness, \cref{tab:eh-math} lists other possible algebraic operations on $x_0$ and $x_1$ enabled by \cref{circ:eh-sum}.  
These operations are obtained by measuring the qubits of the final state $\ket{\psi_\Sigma}$ in other bases.

\begin{table}[htbp]
\caption{\label{tab:eh-math}%
Expectation values for \cref{circ:eh-sum} for projective single-qubit measurement in the \(X\), \(Y\), and \(Z\) bases.}
\centering
\begin{minipage}{0.6\textwidth}
\renewcommand{\arraystretch}{1.5} 
\begin{ruledtabular}
\begin{tabular}{lcc}
\textbf{Measurement Basis}
  & \(\text{EV}[q_0] = \langle\psi_\Sigma|O_p|\psi_\Sigma\rangle\)
  & \(\text{EV}[q_1] = \langle\psi_\Sigma|O_p|\psi_\Pi\rangle\) \\[0.5ex]
\colrule
\qquad\quad\raisebox{0.25em}{\Qcircuit @C=0.5em {& \qw & \measuretab{X}}}~\footnote{Measuring in the \(X\)-basis:\, 
\raisebox{0.25em}{\Qcircuit @C=0.5em {& \qw & \qw & \qw & \qw & \qw & \gate{H} & \measuretab{Z}}}\\[-1.5ex]~}  
  & \(\sqrt{w(1-w)}(x_0 - x_1)\) 
  & \(\sqrt{1-w} \cdot \sqrt{1-x_0^2}\) \\[0.5ex]

\qquad\quad\raisebox{0.25em}{\Qcircuit @C=0.5em {& \qw & \measuretab{Y}}}~\footnote{Measuring in the \(Y\)-basis: 
\raisebox{0.25em}{\Qcircuit @C=0.5em {& \qw & \gate{S^\dagger} & \gate{H} & \measuretab{Z}}}}  
  & \(\sqrt{1-x_0^2} \cdot \sqrt{1-x_1^2}\) 
  & \(\sqrt{w} \cdot \sqrt{1-x_1^2}\) \\[0.5ex]

\qquad\quad\raisebox{0.25em}{\Qcircuit @C=0.5em {& \qw & \measuretab{Z}}}
  & \(w \cdot x_0 + (1-w) \cdot x_1\) 
  & \(x_0 \cdot x_1\) \\ 
\end{tabular}
\end{ruledtabular}
\end{minipage}
\end{table}

\subsection{Linear combinations by concatenating summation circuits}

We now generalize \cref{circ:eh-sum} to compute the linear combination of three real numbers.  
This generalization uses three qubits and two different $\Usum$ gates.
Our objective is to compute the linear combination of 3 real values \( \{x_0,x_1,x_2\} \in [-1, 1]\):
\begin{equation}
    s_3=a_0x_0 +  a_1x_1 +  a_2x_2,
    \label{eq:s3}
\end{equation}
where the coefficients satisfy the following constraints
\begin{align}
    a_0+a_1+a_2 &= 1, & a_i &\in [0,1].
    \label{eq:a123=1}
\end{align}
Strictly speaking, the coefficients $a_i$ must be non-negative. However, we will later introduce a trick involving the insertion of $X$-gates, enabling us to also handle cases with negative coefficients as well.

First, we need to expand the input EVEN encoding to encode 3 real values in the following 3-qubit state
\begin{equation}
    \ket{\phi_3} = \left[ \Ry(\theta_0) \otimes \Ry(\theta_1) \otimes \Ry(\theta_2) \right] \ket{000},
    \label{eq:x3theta}
\end{equation}
where $\theta_i$ are given in \cref{eq:app-theta}.
Intuitively, the concatenation of two $\Usum$ circuits is expect to produce
\begin{equation}
s_3 = w_1\left[w_0x_0+(1-w_0)x_1\right]+(1-w_1)x_2
    =\underbrace{w_1w_0}_{a_0} \cdot\, x_0 + \underbrace{w_1(1-w_0)}_{a_1} \cdot\, x_1 + \underbrace{(1-w_1)}_{a_2} \cdot\, x_2,
    \label{eq:v2w}
\end{equation}
where the two $\Usumsum$-parameters $w_0$ and $w_1$ can be found by solving the set of \cref{eq:s3,eq:v2w}. 
The particular 
case of $ w_0=1/2$, $w_1=2/3$ 
results
in the plain average over 3 numbers:
$(x_0+x_1+x_2)/3$.
The corresponding quantum circuit, shown in \cref{circ:eh-sumsumR}, combines $\Usum(w_0)$, acting on qubits $\{0,1\}$, with $\Usum(w_1)$, acting on qubits $\{0,2\}$.

\begin{circuit}[hbtp]
\centering
\mbox{%
\Qcircuit @C=0.7em @R=1.2em {
& & & & & \lstick{\ket{0}_{\text{anc}}} & \qw  & \gate{H}&   \ctrl{1} & \qw & \qw \\
& \lstick{\ket{0}_0} & \gate{\Ry(\theta_0)} \barrier[0em]{2} & \qw & \qw & \multigate{1}{\Usum(w_0)}& \qw & \qw  & \ctrl{-1} & \qw&\multigate{2}{\Usum(w_1)} & \barrier[0em]{2} \qw & \qw & \measuretab{Z} \\
& \lstick{\ket{0}_1} & \gate{\Ry(\theta_1)} & \qw & \qw & \ghost{\Usum(w_0) } & \qw \\
& \lstick{\ket{0}_2} & \gate{\Ry(\theta_2)} & \qw & \qw & \qw & \qw & \qw & \qw & \qw & \ghost{\Usum(w_1)} & \qw & \qw & \qw \\
&&& \ket{\phi_3} &&&\Usumsum(w_0,w_1) & && &&&\ket{\psi_3}\\
&
{\gategroup{1}{8}{2}{9}{1.2em}{.}}
}}
\caption{\justifying\small Circuit yielding  expectation value being the weighted sum of 3 input values $\text{EV}=w_1\left[w_0x_0+(1-w_0)x_1\right]+(1-w_1)x_2$.}
\label{circ:eh-sumsumR}
\end{circuit}

Measuring qubit $q_0$ in \cref{circ:eh-sumsumR} for the operator $O_{s3} := \sigma_z \otimes I \otimes I$ and state $\ket{\psi_3} = \Usumsum \ket{\phi_3}$, yields the EV
\begin{equation}
    \bra{\psi_3}O_{s3}\ket{\psi_3} = w_1\left[w_0x_0+(1-w_0)x_1\right]+(1-w_1)x_2.
    \label{eq:sum3}
\end{equation}
The use of the parity-flip operator, indicated by the dotted rectangle in  \cref{circ:eh-sumsumR}, is essential to obtain the correct result in \cref{eq:sum3}. Omitting this operator introduces an unwanted term in the output, specifically the additional square root term shown in \cref{eq:sum3-unwanted}:
\begin{equation}
w_1\left[w_0x_0+(1-w_0)x_1\right]+(1-w_1)x_2 
\;+\;\tfrac12\,\sqrt{(-1 + w_1)\,w_1}\,
      \sqrt{1 - x_0^2}\,\sqrt{1 - x_1^2}\,\sqrt{1 - x_2^2}.
      \label{eq:sum3-unwanted}
\end{equation}
Thus, the parity-flip operator in \cref{circ:eh-sumsumR} serves to cancel the additional square roots term.

If the reversibility is not required, the {\it ancilla} qubit can be reset and reused after the CZ gate is applied. Similarly, qubits 1 and 2 can be reset and reused after their respective gate blocks are completed, enabling efficient qubit recycling for more complex \eh\ circuits.

Note that it is possible to obtain a modified version of \cref{circ:eh-sumsumR} to handle a sum with a negative coefficient $a_j<0$. This can be achieved by inserting an $X$-gate before the $\Ry(\theta_j)$ gate for the specific input qubit $q_j$. It is equivalent to flipping the sign of $x_j\rightarrow -x_j$ while using $|a_j|>0$. 
 
\Cref{circ:eh-sumsumR} can be further generalized to compute a weighted sum over $K$ inputs by concatenating $K-1$ $\Usum$ unitaries, interspersed with $K-2$ parity-flip operators consuming $K-2$ ancillas. For any set of desired coefficients $a_i$,  subject to 
\begin{align}
    \sum_i a_i &= 1, & a_i &\in [0,1],
    \label{eq:a12345=1}
\end{align}
it is always possible to determine $K-1$ $\Usum$-parameters $w_j$.
E.g. a plain average over $K$ variables: $\frac{1}{K}\sum_{i=0}^{K-1} x_i$ can be obtained using weights: $w_i=\frac{i+1}{i+2}$ for $i\in[0,\ldots,K-2]$.

\subsection{Polynomial in multiple variables}
\label{secSM:multi-var}

The \ehp\  can  be extended to compute higher degree polynomials involving multiple variables.
Consider a polynomial $p(x_1, x_2,\dots,x_n)$ of degree $d$, which can be expressed as:
\begin{align}
    \label{eq:P(xyz)}
p(x_1, x_2,\dots,x_n) = \sum_{i=0}^{d}a_ix_1^{i_1}x_2^{i_2}\dots x_n^{i_n},
\end{align}
where $a_i$ are real coefficients, and each term $x_1^{i_1}x_2^{i_2}\dots x_n^{i_n}$ represents a product of the input variables raised to different powers, with $i_1+i_2+\dots+i_n \leq d$.
All input values $x_i$ and coefficients $a_i$ are constrained to the range $[-1,1]$, ensuring that the polynomial terms are bounded and computable within the limits of the \ehp.

\section{Computation of Mean Squared Error with \ehptabs}
\label{secSM:MSE}

The mean squared error (MSE) is widely used in machine learning to evaluate the performance of regression models. 
It measures the average squared differences between the true values $y_i$ and predicted values  $x_i$, helping assess model accuracy and guiding optimization processes.
The MSE is defined as:
\begin{align}
\text{MSE}(\mathbf{x}, \mathbf{y}) = \frac{1}{N} \sum_{i=0}^{N-1}\left(x_i - y_i\right)^2,
\end{align}
where $N$ is the number of predicted values.

\begin{circuit}[hbtp]
    \centering
    \mbox{%
    \Qcircuit @C=0.5em @R=0.2em @!R{\\
    \lstick{\ket{0}_{anc_0}} & \qw \barrier[-1em]{18} & \qw  & \qw & \gate{H}& \qw & \qw & \qw & \qw & \qw & \qw & \qw & \qw & \ctrl{3} & \qw & \qw
    \\ 
    \lstick{\textcolor{blue}{x_0  \rightarrow }~~\ket{0}_{d_0}} & \gate{\Ry(\alpha_0)} & \gate{X} & \qw &  \multigate{1}{\Sigma_{\frac{1}{2}}}_<<{1} & 
    \\ 
    \lstick{\textcolor{blue}{y_0  \rightarrow }~~\ket{0}_{d_1}} & \gate{\Ry(\beta_0)} & \qw & \qw & \ghost{\Sigma_{\frac{1}{2}}}_<<{0} & \multigate{1}{\Pi}  &
    \\ 
    \lstick{\textcolor{blue}{x_0  \rightarrow }~~\ket{0}_{d_2}} & \gate{\Ry(\alpha_0)} & \gate{X} & \qw &  \multigate{1}{\Sigma_{\frac{1}{2}}}_<<{0} & \ghost{\Pi}  & \qw & \qw & \qw & \ustick{\frac{(y_0-x_0)^2}{4}} \qw & \qw & \qw & \qw & \control \qw & \qw & \multigate{4}{\Sigma_{\frac{1}{2}}}_<<{1} &
    \\ 
    \lstick{\textcolor{blue}{y_0  \rightarrow }~~\ket{0}_{d_3}} & \gate{\Ry(\beta_0)} & \qw & \qw & \ghost{\Sigma_{\frac{1}{2}}}_<<{1} & & & & & & & & & & & \nghost{\Sigma_{\frac{1}{2}}} &
    \\ 
    \lstick{\textcolor{blue}{x_1  \rightarrow }~~\ket{0}_{d_4}} & \gate{\Ry(\alpha_1)} & \gate{X} & \qw &  \multigate{1}{\Sigma_{\frac{1}{2}}}_<<{1} & & & & & & & & & & & \nghost{\Sigma_{\frac{1}{2}}} &
    \\ 
    \lstick{\textcolor{blue}{y_1  \rightarrow }~~\ket{0}_{d_5}} & \gate{\Ry(\beta_1)} & \qw & \qw & \ghost{\Sigma_{\frac{1}{2}}}_<<{0} & \multigate{1}{\Pi} & & & & & & & & & & \nghost{\Sigma_{\frac{1}{2}}} &
    \\ 
    \lstick{\textcolor{blue}{... }~~\ket{0}_{d_6}} & \gate{\Ry(\alpha_1)} & \gate{X} & \qw &  \multigate{1}{\Sigma_{\frac{1}{2}}}_<<{0} & \ghost{\Pi}  & \qw & \qw & \qw & \ustick{\frac{(y_1-x_1)^2}{4}} \qw & \qw & \qw & \qw & \qw & \qw & \ghost{\Sigma_{\frac{1}{2}}}_<<{0} & \qw & \multigate{8}{\Sigma_{\frac{1}{2}}}_<<{1} &
    \\ 
    \lstick{\textcolor{blue}{... }~~\ket{0}_{d_7}} & \gate{\Ry(\beta_1)} & \qw & \qw & \ghost{\Sigma_{\frac{1}{2}}}_<<{1} &
    \\ 
    \lstick{\textcolor{blue}{... }~~\ket{0}_{d_8}} & \gate{\Ry(\alpha_2)} & \gate{X} & \qw &  \multigate{1}{\Sigma_{\frac{1}{2}}}_<<{1} &
    \\ 
    \lstick{\textcolor{blue}{... }~~\ket{0}_{d_9}} & \gate{\Ry(\beta_2)} & \qw & \qw & \ghost{\Sigma_{\frac{1}{2}}}_<<{0} & \multigate{1}{\Pi} &
    \\ 
    \lstick{\textcolor{blue}{... }~~\ket{0}_{d_{10}}} & \gate{\Ry(\alpha_2)} & \gate{X} & \qw &  \multigate{1}{\Sigma_{\frac{1}{2}}}_<<{0} & \ghost{\Pi}  & \qw & \qw & \qw & \ustick{\frac{(y_2-x_2)^2}{4}} \qw & \qw & \qw & \qw & \qw & \qw &  \multigate{4}{\Sigma_{\frac{1}{2}}}_<<{1} &
    \\ 
    \lstick{\textcolor{blue}{... }~~\ket{0}_{d_{11}}} & \gate{\Ry(\beta_2)} & \qw & \qw & \ghost{\Sigma_{\frac{1}{2}}}_<<{1} & & & & & & & & & & & \nghost{\Sigma_{\frac{1}{2}}} &
    \\ 
    \lstick{\textcolor{blue}{... }~~\ket{0}_{d_{12}}} & \gate{\Ry(\alpha_3)} & \gate{X} & \qw &  \multigate{1}{\Sigma_{\frac{1}{2}}}_<<{1} & & & & & & & & & & & \nghost{\Sigma_{\frac{1}{2}}} &
    \\ 
    \lstick{\textcolor{blue}{... }~~\ket{0}_{d_{13}}} & \gate{\Ry(\beta_3)} & \qw & \qw & \ghost{\Sigma_{\frac{1}{2}}}_<<{0} & \multigate{1}{\Pi} & & & & & & & & & & & \nghost{\Sigma_{\frac{1}{2}}} &
    \\ 
    \lstick{\textcolor{blue}{x_3  \rightarrow }~~\ket{0}_{d_{14}}} & \gate{\Ry(\alpha_3)} & \gate{X} & \qw &   \multigate{1}{\Sigma_{\frac{1}{2}}}_<<{0} & \ghost{\Pi}  & \qw & \qw & \qw & \ustick{\frac{(y_3-x_3)^2}{4}} \qw & \qw & \qw & \qw & \ctrl{2} & \qw & \ghost{\Sigma_{\frac{1}{2}}}_<<{0} & \ctrl{3} & \ghost{\Sigma_{\frac{1}{2}}}_<<{0} & \qw & \qw & \qw & \qw &  \measuretab{Z} & \rstick{\textcolor{blue}{\rightarrow \frac{1}{16}\sum_{i=0}^{3} (y_i-x_i)^2}}
    \\ 
    \lstick{\textcolor{blue}{y_3  \rightarrow }~~\ket{0}_{d_{15}}} & \gate{\Ry(\beta_3)} & \qw & \qw & \ghost{\Sigma_{\frac{1}{2}}}_<<{1} &
    \\ 
    \lstick{\ket{0}_{anc_1}} & \qw & \qw & \qw & \gate{H} & \qw & \qw & \qw & \qw & \qw & \qw & \qw & \qw & \control \qw & \qw & \qw
    \\ 
    \lstick{\ket{0}_{anc_2}} & \qw & \qw &  \qw & \gate{H} & \qw & \qw & \qw & \qw & \qw & \qw & \qw & \qw & \qw & \qw & \qw & \control \qw & \qw &  &  &  &
    \\ 
  }}
\caption{\justifying\small \eh\ computation of  mean squared error (MSE) on two vectors of real values $x_i,y_i$ of length 4. The explicit implementation of unitaries $\sum_\frac{1}{2},\Pi$   are defined in Fig.~1 in the main text.
}
\label{fig:eh-mse}
\end{circuit}

The \ehp\  can be used to compute the MSE on a QPU as illustrated in~\cref{fig:eh-mse} for $N=4$, pairs  of real values, $x_i,y_i$.
For this MSE implementation, the total \CNOT\ count is 29 and the \CNOT\ depth is 9. 
The computation proceeds as follows:
\begin{itemize}
    \item First, the EVEN encoding is used to store 4 copies of both inputs $x_i,y_y$ on 16 qubits, labeled $d_0,...,d_{15}$.  $R_y$ rotation angles are: $ \alpha_i=\arccos{x_i}$,  $ \beta_i=\arccos{y_i}$ .
    \item The difference between each pair of values $(y_i-x_i)$ is calculated twice. We negate all y-inputs so despite we use the sum operator with weight $w=\nicefrac{1}{2}$ the results is the difference.
    \item The two independent copies of the differences are multiplied, yielding the terms $\frac{1}{4}(y_i-x_i)^2$.
    \item Finally, those squared terms are averaged, resulting with the MSE which can be measured as the expectation value of one qubit (labeled $d_{14}$ in~\cref{fig:eh-mse}). 
\end{itemize}
The four copies of each input are necessary, because the {\it no-cloning} theorem prohibits using same quantum state twice, what is needed to square the values of the differences.
Notice that the final output is $\text{MSE}/4$ 
which is a characteristic feature of the \ehp, due to how the square of the difference is computed. This attenuation factor of 4 remains unchanged as the number of input variables 
($N$) increases. For the purpose of minimization, a constant scaling of the loss does not pose a problem. However, in practice, more shots are required to achieve a given statistical accuracy for the MSE.

In general, the quantum resources required to compute MSE for $N$ pairs of real values $x_i, y_i$ using the \ehp\ and serial sum  implementation are as follows:  
(i) \(4N+\mathcal{O}(\log N)\) qubits,  
(ii) \(5N+ \mathcal{O}(\log N)\) \CNOT\ gates,  
(iii) \(\mathcal{O}(\log N)\) \CNOT\ depth,
(iv) one qubit measured at the end.  

The potential application of vectorized MSE computation using the \ehp\  with the \qcrank\ encoding~\cite{qbart} include, for example, the computation of batch-average loss for QML applications.
The edge detection for gray-scale images is another example.
It is one of the simplest and most useful image-processing operations is edge detection, which involves
computing the squared gradient between pairs of pixels and applying a threshold on it. The
threshold can be easily adjusted to account for reduced gradient magnitudes caused by limited gate
fidelity on the NISQ hardware. This approach may require more shots to achieve comparable precision on hardware
versus an ideal simulator.

\section{Numerical simulations}
\label{secSM:simulations}

To demonstrate the flexibility of the \ehp, we selected three common functions listed in~\cref{tab:sim}. 
Each function was approximated by a polynomial of degree 5 or 6, depending on its parity, through classical pre-processing. These polynomials were then implemented as \eh\ quantum circuits by naturally extending the pattern shown in \cref{fig:eh-poly_4-rev} in the main text to the required degree \(d\).

\begin{table}[h!]
\caption{\label{tab:sim}Polynomial approximations of selected functions for testing the performance of the reversible \ehp\ in ideal simulations.}
\begin{ruledtabular}
\begin{tabular}{lccccc}
\textrm{target} &
\textrm{\cref{fig:sim}} &
\textrm{poly} &
\textrm{num} &
\textrm{num}&
\textrm{used} \\ 
\textrm{function} &
\textrm{panel} &
\textrm{degree} &
\textrm{CNOT} &
\textrm{qubits}&
\textrm{shots\footnote{number of shots per $x$-data point}} \\
\colrule
$\exp(2x)$    & a & 5 & 23 & 15 & 2e5\\
$\arctan(5x)$ & b & 5 & 23 & 15 & 1e7\\
$\exp(-9x^2)$ & c & 6 & 28 & 18 & 8e7\\
\end{tabular}
\end{ruledtabular}
\end{table}

\begin{figure*}[hbtp]
\centering
\includegraphics[width=.99\linewidth]{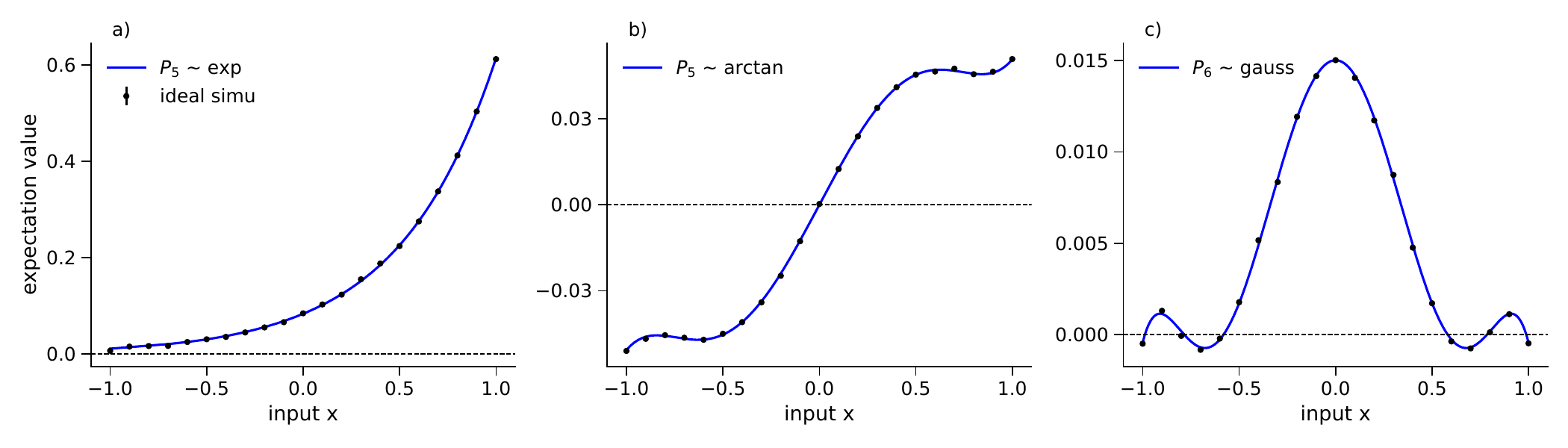}
   \caption{Polynomial approximations of the selected functions listed in \cref{tab:sim} are accurately reproduced by ideal Qiskit simulations. The solid curves represent the ground truth, while the data points, shown with statistical error bars, were obtained from a shot-based simulator using the number of shots specified in the last column of the table.}
\label{fig:sim}
\end{figure*}

Those 3 quantum circuits were simulated using the shot-based ideal Qiskit simulator for a grid of 21 equally spaced input values \(x \in [-1, 1]\). As shown in~\cref{fig:sim}, the simulated Qiskit results closely match the analytical values of each polynomial (solid line), demonstrating strong agreement and confirming the accuracy of our approach.

\section{Hardware performance optimization}
\label{secSM:hardware}

The results of executing \ehp\ on IBM hardware, shown in \cref{fig:ibm} in the main text, represent the best performance observed on the day the experiments were conducted (May 2025). For those main results, we employed randomized compilation and optimized qubit mapping; however, we did not utilize dynamic decoupling or zero noise extrapolation.
In this section, we present additional exploratory experiments on the IBM hardware that justify our choices of error mitigation tools.

\subsection{Impact of randomized compilation and dynamic decoupling}

Randomized compilation (RC), also known as Pauli twirling, mitigates coherent errors by converting them into stochastic noise. Dynamic decoupling (DD) inserts a carefully timed sequence of pulses to each qubit timeline, aiming to cancel a portion of the surrounding noise. Potentially, combining these techniques yields more reliable and statistically robust results on NISQ devices~\cite{8lxc-lvv1}.

\begin{figure*}[hbtp]
\centering
\includegraphics[width=.99\linewidth]{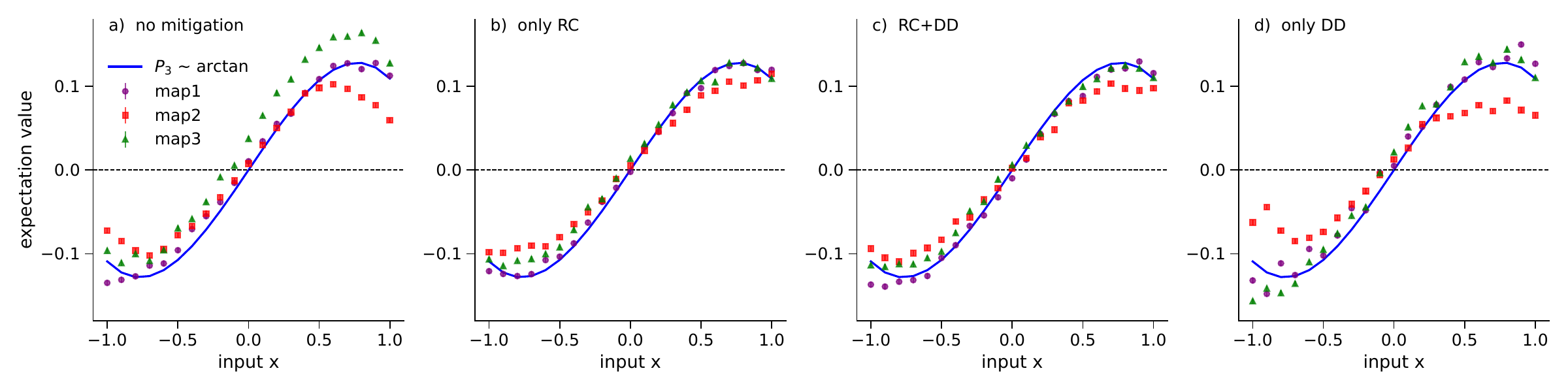}
\caption{Impact of randomized compilation (RC) and/or dynamic decoupling (DD) on MSE of $P_3(x)\sim \arctan(x)$ measured on IBM Marrakesh for the same circuit transpiled onto three sets of qubits, labeled as map1, 2, 3. See~\cref{tab:RC} for the details, including MSE of each measurement.}
\label{fig:RC}
\end{figure*}

For one particular choice of $P_d(x)\sim \arctan(x)$ polynomial and using one of the best-performing QPU on the day the research was conducted—namely IBM Marrakesh—we examined the impact of enabling RC, DD, and qubit mappings
 on the fidelity of the outcome. This was done for three 
\textit{a priori}
equivalent qubit layouts, each selected by the standard IBM transpiler using different random seeds. All 12 jobs were executed within 2 hours wall-time.
The MSE between ground truth polynomial value and the measurement was the infidelity metric of choice. The results are shown in~\cref{tab:RC} and in~\cref{fig:RC}.

\begin{table}[h!]
\caption{MSE for  $P_d(x)\sim \arctan(x)$ polynomial transpiled to different subset of qubits and executed on IBM Marrakesh with and without randomized compilation. MSE was  multiplied by 1e4 for clarity. 
}
\label{tab:RC}
\begin{ruledtabular}
\begin{tabular}{cccccc}
mapping & qubits & MSE  raw  & MSE only RC & MSE w/ RC+DD& MSE only DD\\
\hline
map1 & 44, 42, 56, 43, 45, 46, 47, 37, 57        &  $1.1 \pm  0.3$  &  $0.4 \pm  0.1$  & $1.1 \pm  0.3$  & $2.4 \pm  0.7$ \\
map2 & 9, 27, 28, 17, 7, 5, 6, 8, 4                & $7.1 \pm  2.2$   & $4.4 \pm  1.4$     & $4.1 \pm  1.3$    & $15.6 \pm  4.8$  \\
map3 & 107, 109, 87, 97, 106, 105, 104, 108, 117 & $10.8 \pm  3.3$ & $1.7 \pm  0.5$ & $1.1 \pm  0.3$   & $3.5 \pm  1.1$   \\
\end{tabular}
\end{ruledtabular}
\end{table}

Due to the daily variability in the calibration of the Marrakesh QPU, the specific choice of qubits lacks intrinsic significance. We list the three sets of 9 qubit mappings here for illustrative purposes only. Randomized compilation  and dynamical decoupling  were enabled individually and in combination using the Qiskit SamplerOptions(). For RC, we selected twirling for both gates and measurements, setting the number of randomizations to 60. For DD, we used the 'XX' sequence type, with slack distribution set to 'middle' and scheduling configured as 'ALAP'.
To keep statistical error constant we always used the same total number of shots per circuit.
 
The improvement due to enabled randomized compilation was significant, as was the manual selection of the optimal qubit mapping, as shown in~\cref{fig:RC}b. In contrast, the effect of dynamical decoupling (DD) was only marginal, resulting in a change in mean squared error (MSE) within one standard deviation (see panel c and d). Therefore, applying Occam's razor principle, we decided to omit DD. While it is possible to combine RC with the zero noise extrapolation (ZNE) technique for enhanced noise mitigation, doing so would require submitting several times more jobs and is beyond the scope (and budget) of this paper.

Based on this study, we have concluded that all IBM hardware results in this work will be executed using both RC and optimal qubit mapping selection, while omitting DD and ZNE.

\subsection{Selection of qubit mapping on hardware}

To confirm and better justify our decision to manually select the optimal qubit mapping, using the Qiskit command \texttt{transpile(qc, backend,optimization\_level=3, random\_seed=42)}, we conducted all experiments presented in the main paper twice, using two different mappings, and chose the one with the observed smaller mean squared error (MSE). We enabling randomized compilation for all those experiments, as discussed in the previous section. This section provides all results and  discussion of the procedure we employed.

For the given input circuit, the \eh\ circuits require only a small subset of the qubits available on IBM QPUs. The Qiskit transpiler automatically selects a locally optimal qubit mapping based on the current calibration of the target QPU. While the transpiler’s heuristic generally yields a good layout, it does not guarantee that a global optimum is found.
By varying the transpiler’s random seed, we explored different locally optimal mappings. The only practical way to assess the quality of those different mappings is to execute the circuit on each and compare the resulting MSE to select the best mapping.

\begin{figure*}[hbtp]
\centering
\includegraphics[width=.99\linewidth]{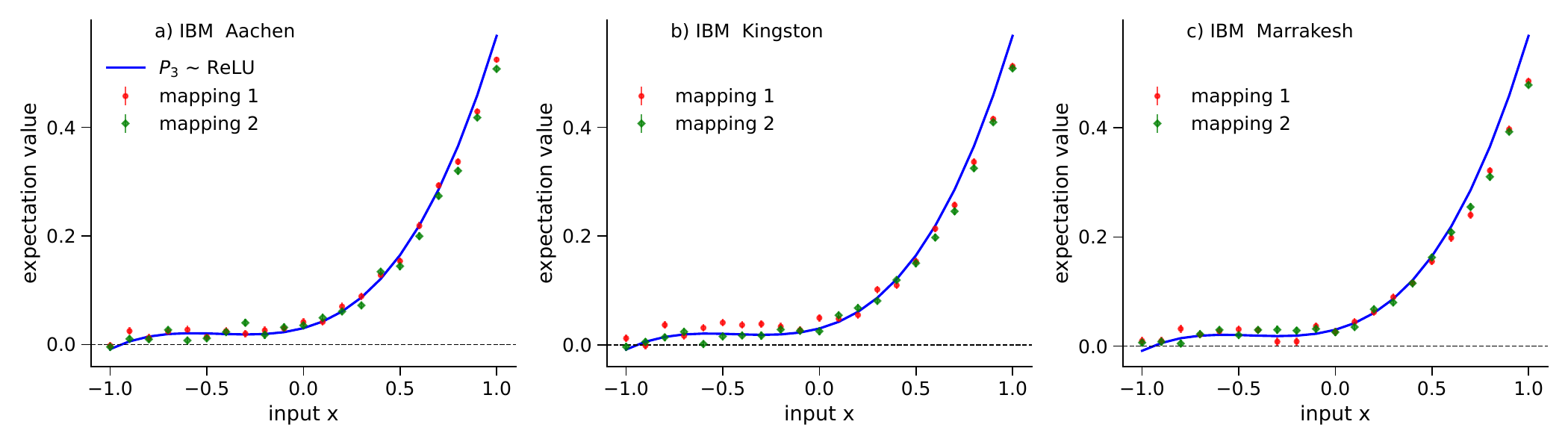}
\caption{Dependence of $P_3(x)\sim $ReLU MSE fidelity on the choice of QPU and qubit subset. The randomized compilation was always enabled.}
\label{fig:best-relu}
\end{figure*}

\begin{figure*}[hbtp]
\centering
\includegraphics[width=.99\linewidth]{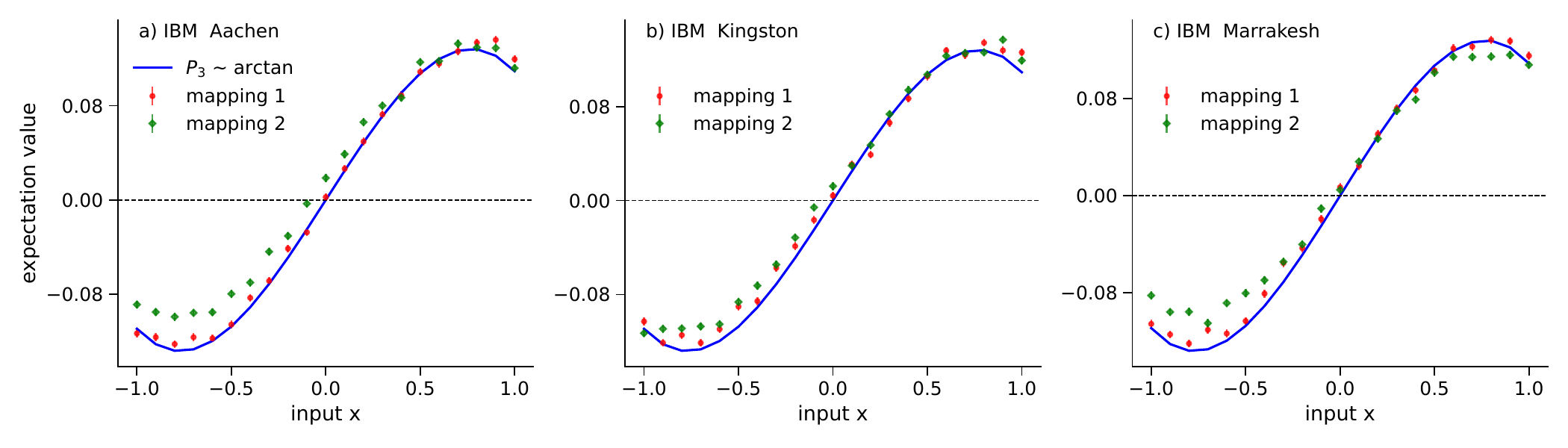}
\caption{Dependence of $P_3(x)\sim \arctan(x)$ MSE fidelity on the choice of QPU and qubit subset.}
\label{fig:best-atan}
\end{figure*}

\begin{figure*}[hbtp]
\centering
\includegraphics[width=.99\linewidth]{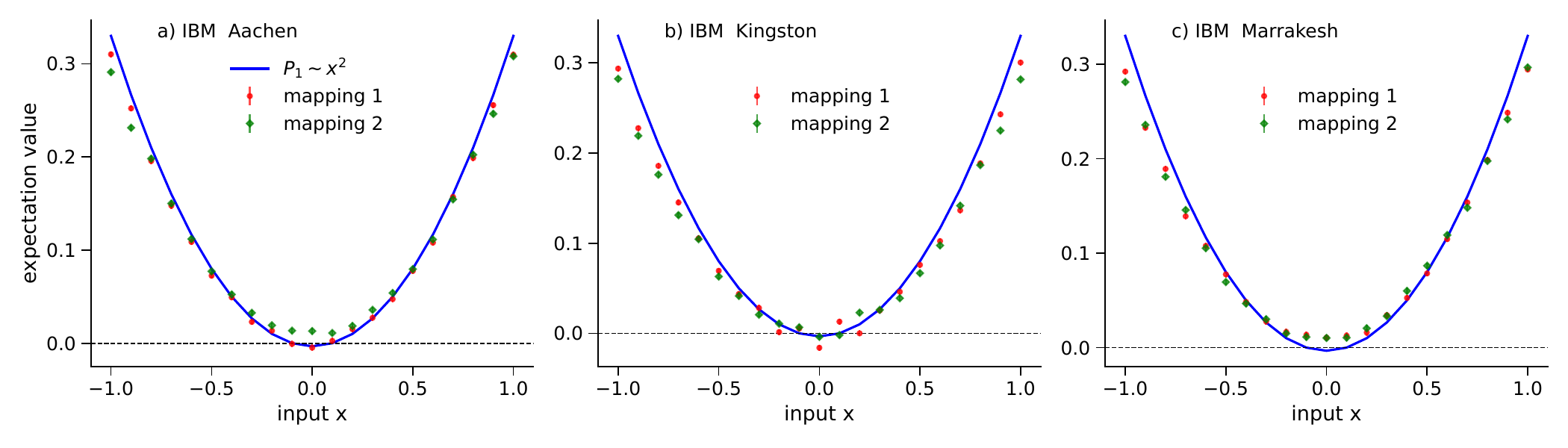}
\caption{Dependence of $P_2(x)\sim x^2$ MSE fidelity on the choice of QPU and qubit subset.}
\label{fig:best-x2}
\end{figure*}

We evaluated two mappings for each of the three polynomial types across all three QPUs used in this work.
\cref{tab:best} lists the MSE values for all 18 tested configurations. Randomized compilation was always enabled. \Cref{fig:best-relu,fig:best-atan,fig:best-x2} present the complete set of obtained results, grouped by polynomial type.  We show the better-performing mapping (i.e., the one with lower MSE) for each function–QPU pair, highlighted in red.
The better results are shown in \cref{fig:ibm} in the main text.

\begin{table}[h!]
\caption{MSE  dependence on the choice of 3 QPUs and 2 subsets of qubits used to execute the circuit with enabled randomized compilation.  MSE was  multiplied by 1e4 for clarity.
}
\label{tab:best}
\begin{ruledtabular}
\begin{tabular}{ccccc}
QPU  & phys qubits \footnote{Mappings are different for every QPU and every polynomial} \footnote{Mappings are ordered from the best  MSE to the worst} &$P_3\sim$ ReLU & $P_3\sim $  arctan & $P_2 \sim  x^2$\\
\hline
Aachen &mapping 1 &  $2.2 \pm  0.7$&  $0.3 \pm  0.1$  & $0.9 \pm  0.3$\\
        &mapping 2  &  $4.7 \pm  1.4$  &  $3.5 \pm  1.1$    &  $2.3 \pm  0.7$  \\
  \hline
Kingston  &mapping 1 & $4.8 \pm  1.5$ &  $0.8 \pm  0.2$  &  $3.4 \pm  1.0$ \\
          &mapping 2 &  $5.1 \pm  1.6$  &   $1.6 \pm  0.5$   &$6.1 \pm  1.9$  \\
\hline
Marrakesh  &mapping 1  &$7.9 \pm  2.4$&   $0.5 \pm  0.1$ & $2.8 \pm  0.9$\\
         &mapping 2   & $8.4 \pm  2.6$ &    $3.0 \pm  0.9$   & $3.5 \pm  1.1$\\
\hline
    shots     & & 2e4 & 9e4 & 9e4\\
\end{tabular}
\end{ruledtabular}
\end{table}

\section{Shallow depth \ehtabs\ polynomial}
\label{secSM:shallow}

The main text shows two alternative implementations of $P_d(x)$. \Cref{fig:eh-poly_4-rev} shows a reversible version requiring $3d$ qubits, and a non-reversible version requiring $d$ qubits, shown in \cref{fig:eh-poly_4-irrev}. The circuits for both of these versions have the same CNOT depth of $5d - 2$.

\cref{fig:eh-poly_6_2} shows yet another  $P_d(x)$ implementation whose CNOT gate depth scales as $\mathcal{O}(\log d)$, at the expense of storing the input in more qubits in comparison to other versions.
From its construction, it is clear that $\log(d)$ depth of multiplicity operators, $\Pi$, is required to form all powers of $x$, up to $x^d$. Similarly, $\log(d)$ depth of sum operators, $\sum$, is required to add $d+1$ terms for $P_d(x)$. Since it takes 1 CNOT for every $\Pi$ and 2 CNOTs for every $\sum$, the total CNOT depth of $P_d(x)$ is of $O(\log(d))$.

Naturally, one needs to compensate for the shallow circuit with more qubits carrying more copies of the input value $x$, to allow for the quick build-up of all powers of $x$ without reusing the same qubits, to not violate non cloning theorem. 

\begin{circuit*}[htbp]
    \centering
    \hspace*{4.25em}%
    \scalebox{0.79}{
    \Qcircuit @C=0.5em @R=0.2em @!R{\\
    \\ 
    \lstick{\textcolor{blue}{x  \rightarrow }~~\ket{0}_{d_0}} & \gate{\Ry(\theta)} & \qw \barrier[-2.3em]{19} & \multigate{1}{\Pi} & \qw \barrier[-1.5em]{19} & \qw  & \qw & \qw & \qw & \qw & \qw  \barrier[-2.9em]{19} & \qw & \qw & \qw & \ustick{x} \qw & \qw & \qw \barrier[-.5em]{19} & \qw & \qw & \qw & \qw & \qw & \qw & \qw & \qw & \qw & \qw & \qw & \multigate{15}{\Pi}_<<{0} & \qw &  &  &  &  &  &  &  &  &  &  &  &  &  &  &  &  &  &  &  &  &  &  &
    \\ 
    \lstick{\textcolor{blue}{...}~~\ket{0}_{d_1}} & \gate{\Ry(\theta)} & \qw & \ghost{\Pi} & \qw & \multigate{2}{\Pi}_<<{0} & \qw & \qw & \qw & \qw & \qw & \qw & \qw & \qw & \ustick{x^2} \qw & \qw & \qw & \qw & \qw & \qw & \qw & \qw & \qw & \qw & \qw & \qw & \multigate{13}{\Pi}_<<{0} & \qw & \nghost{\Pi} &  &  &  &  &  &  &  &  &  &  &  &  &  &  &  &  &  &  &  &  &  &  &  &
    \\ 
    \lstick{\textcolor{blue}{...}~~\ket{0}_{d_2}} & \gate{\Ry(\theta)} & \qw & \multigate{1}{\Pi} & \qw & \ghost{\Pi}& \qw & \multigate{3}{\Pi}_<<{0} & \qw &  &   &   &   &   &   &   &   &  &  &  &  &  &  &  &  &  & \nghost{\Pi} &  & \nghost{\Pi} &  &  &  &  &  &  &  &  &  &  &  &  &  &  &  &  &  &  &  &  &  &
    \\ 
    \lstick{\textcolor{blue}{...}~~\ket{0}_{d_3}} & \gate{\Ry(\theta)} & \qw & \ghost{\Pi} & \qw & \ghost{\Pi}_<<{1} & \qw & \ghost{\Pi} & \qw & \qw & \qw & \qw & \multigate{6}{\Pi}_<<{0} & \qw &\ustick{x^4}  \qw & \qw & \qw & \qw & \qw &  \qw & \qw & \qw & \multigate{9}{\Pi}_<<{0} & \qw &  &  & \nghost{\Pi} &  & \nghost{\Pi} &   &  &  &  &  &  &  &  &  &  &  &  &  &  &  &  &  &  &  &  &  &  &
    \\ 
    \lstick{\textcolor{blue}{...}~~\ket{0}_{d_4}} & \gate{\Ry(\theta)} & \qw & \multigate{1}{\Pi} & \qw &  &  & \nghost{\Pi} &  &  &  &  & \nghost{\Pi} &  &  &  &  &  &  &  &  &  & \nghost{\Pi} &  &  &  & \nghost{\Pi} &  & \nghost{\Pi} &  &  &  &  &  &  &  &  &  &  &  &  &  &  &  &  &  &  &  &  &  &
    \\ 
    \lstick{\textcolor{blue}{...}~~\ket{0}_{d_5}} & \gate{\Ry(\theta)} & \qw & \ghost{\Pi} & \qw & \qw & \qw & \ghost{\Pi}_<<{1} & \qw & \qw & \multigate{2}{\Pi}_<<{0} & \qw & \ghost{\Pi} & \qw & \ustick{x^3} \qw & \qw & \qw & \qw & \qw & \qw & \qw & \qw & \ghost{\Pi }& \qw & \multigate{8}{\Pi}_<<{0} &\qw  & \nghost{\Pi} &  & \nghost{\Pi} &  &  &  &  &  & &  &  &  &  &  &  &  &  &  &  &  &  &  &  &  &  &  &
    \\ 
    \lstick{\textcolor{blue}{...}~~\ket{0}_{d_6}} & \gate{\Ry(\theta)} & \qw & \multigate{1}{\Pi} & \qw &  &  &   & &  & \nghost{\Pi} &  & \nghost{\Pi} &  &  &  &  &  &  &  &  &  & \nghost{\Pi} &  & \nghost{\Pi} &  & \nghost{\Pi} &  & \nghost{\Pi} &  &  &  &  &  &  &  &  &  &  &  &  &  &  &  &  &  &  &  &  &  &  &  &
    \\ 
    \lstick{\textcolor{blue}{...}~~\ket{0}_{d_7}} & \gate{\Ry(\theta)} & \qw & \ghost{\Pi} & \qw & \qw & \qw & \qw & \qw &  \qw &\ghost{\Pi}_<<{1} & \qw & \ghost{\Pi} & \qw & \ustick{x^5} \qw & \qw & \qw & \qw & \qw &  \qw & \multigate{4}{\Pi}_<<{0} & \qw & \nghost{\Pi} &  & \nghost{\Pi} &  & \nghost{\Pi }&  & \nghost{\Pi} &  &  &  &  &  &  &  &  &  &  &  &  &  &  &  &  &  &  &  &  &  &  &  &
    \\ 
    \lstick{\textcolor{blue}{...}~~\ket{0}_{d_8}} & \gate{\Ry(\theta)} & \qw & \multigate{1}{\Pi} & \qw &  &  &  &  &  &  &  & \nghost{\Pi} &  &  &  &  &  &  &  & \nghost{\Pi} &  & \nghost{\Pi} &  & \nghost{\Pi} &  & \nghost{\Pi} &  & \nghost{\Pi} &  &  &  &  &  &  &  &  &  &  &  &  &  &  &  &  &  &  &  &  &  &  &  &
    \\ 
    \lstick{\textcolor{blue}{x  \rightarrow }~~\ket{0}_{d_9}} & \gate{\Ry(\theta)} & \qw & \ghost{\Pi} & \qw & \qw & \qw & \qw & \qw & \qw & \qw & \qw & \ghost{\Pi}_<<{1} & \qw & \ustick{x^6} \qw & \qw & \qw & \qw & \multigate{1}{\Pi}_<<{0} & \qw & \nghost{\Pi} &  & \nghost{\Pi} &  & \nghost{\Pi} &  & \nghost{\Pi} &  & \nghost{\Pi} &  &  &  &  &  &  &  &  &  &  &  &  &  &  &  &  &  &  &  &  &  &  &
    \\ 
    \lstick{\textcolor{blue}{a_6 \rightarrow }~~\ket{0}_{a_6}} & \qw & \qw & \gate{\Ry(\varphi_6)} & \qw & \qw & \ustick{a_6} \qw & \qw & \qw & \qw & \qw & \qw & \qw & \qw & \qw & \qw & \qw & \qw & \ghost{\Pi}_<<{1}  & \qw & \ghost{\Pi} & \qw & \ghost{\Pi} & \qw & \ghost{\Pi} & \qw & \ghost{\Pi} & \qw & \ghost{\Pi}& \qw & \ustick{a_6x^6} \qw & \qw & \qw \barrier[.3em]{9} & \qw & \qw & \multigate{1}{\Sigma \frac{1}{2}} & \qw \barrier[-1.1em]{9}& \qw  & \qw & \qw & \qw & \qw \barrier[0.2em]{9} & \qw & \qw & \multigate{2}{\Sigma \frac{1}{2}} \barrier[0.1em]{9} & \qw & \qw & \qw \barrier[-0.1em]{9} & \qw & \multigate{4}{\Sigma \frac{4}{7}} & \qw \barrier[-1.3em]{9} &  \measuretab{Z}
    \\ 
    \lstick{\textcolor{blue}{a_5 \rightarrow }~~\ket{0}_{a_5}} & \qw & \qw & \gate{\Ry(\varphi_5)} & \qw & \qw & \ustick{a_5} \qw & \qw & \qw & \qw & \qw & \qw & \qw & \qw & \qw & \qw & \qw & \qw & \qw & \qw & \ghost{\Pi}_<<{1}  & \qw & \ghost{\Pi} & \qw & \ghost{\Pi} & \qw & \ghost{\Pi}& \qw & \ghost{\Pi} & \qw & \ustick{a_5x^5} \qw & \qw & \qw & \qw & \qw & \ghost{\Sigma \frac{1}{2}} & \qw &  &  &  &  &  &  &  & \nghost{\Sigma \frac{1}{2}} &  &  &  &  & \nghost{\Sigma \frac{4}{7}} &  &
    \\ 
    \lstick{\textcolor{blue}{a_4 \rightarrow }~~\ket{0}_{a_4}} & \qw & \qw & \gate{\Ry(\varphi_4)} & \qw & \qw & \ustick{a_4} \qw & \qw & \qw & \qw & \qw & \qw & \qw & \qw & \qw & \qw & \qw & \qw & \qw & \qw & \qw & \qw & \ghost{\Pi}_<<{1} & \qw & \ghost{\Pi} & \qw & \ghost{\Pi} & \qw & \ghost{\Pi} & \qw &\ustick{a_4x^4}  \qw & \qw & \qw & \qw & \qw &  \multigate{1}{\Sigma \frac{1}{2}}  & \qw & \qw & \qw& \qw & \ctrl{6}  & \qw & \qw & \qw & \ghost{\Sigma \frac{1}{2}} &  \qw &  &  &  & \nghost{\Sigma \frac{4}{7}} &  &
    \\ 
    \lstick{\textcolor{blue}{a_3 \rightarrow }~~\ket{0}_{a_3}} & \qw & \qw & \gate{\Ry(\varphi_3)} & \qw & \qw & \ustick{a_3} \qw & \qw & \qw & \qw & \qw & \qw & \qw & \qw & \qw & \qw & \qw & \qw & \qw & \qw & \qw & \qw & \qw & \qw & \ghost{\Pi}_<<{1} & \qw & \ghost{\Pi} & \qw & \ghost{\Pi} & \qw & \ustick{a_3x^3} \qw & \qw & \qw & \qw & \qw & \ghost{\Sigma \frac{1}{2}} & \qw &  &  &  &  &  &  &  &  &  &  &  &  & \nghost{\Sigma \frac{4}{7}} &  &
    \\ 
    \lstick{\textcolor{blue}{a_2 \rightarrow }~~\ket{0}_{a_2} }& \qw & \qw & \gate{\Ry(\varphi_2)} & \qw & \qw & \ustick{a_2} \qw & \qw & \qw & \qw & \qw & \qw & \qw & \qw & \qw & \qw & \qw & \qw & \qw & \qw & \qw & \qw & \qw & \qw & \qw & \qw & \ghost{\Pi}_<<{1} & \qw & \ghost{\Pi} & \qw & \ustick{a_2x^2} \qw & \qw & \qw & \qw & \qw &  \multigate{1}{\Sigma \frac{1}{2}}  & \qw & \qw & \qw & \qw & \qw & \qw  & \qw & \qw & \multigate{2}{\Sigma \frac{2}{3}} & \qw & \qw  & \ctrl{5} & \qw& \ghost{\Sigma \frac{4}{7}} & \qw &
    \\ 
    \lstick{\textcolor{blue}{a_1 \rightarrow }~~\ket{0}_{a_1}} & \qw & \qw & \gate{\Ry(\varphi_1)} & \qw & \qw & \ustick{a_1} \qw & \qw & \qw & \qw & \qw & \qw & \qw & \qw & \qw & \qw & \qw & \qw & \qw & \qw & \qw & \qw & \qw & \qw & \qw & \qw & \qw & \qw & \ghost{\Pi}_<<{1} & \qw & \ustick{a_1x} \qw & \qw & \qw & \qw & \qw & \ghost{\Sigma \frac{1}{2}} & \qw &  &  &  &  &  &  &  & \nghost{\Sigma \frac{2}{3}} &  &  &  &  &  &  &
    \\ 
    \lstick{\textcolor{blue}{a_0 \rightarrow }~~\ket{0}_{a_0}} & \qw & \qw & \gate{\Ry(\varphi_0)} & \qw & \qw & \ustick{a_0} \qw & \qw & \qw & \qw & \qw & \qw & \qw & \qw & \qw & \qw & \qw & \qw & \qw & \qw & \qw & \qw & \qw & \qw & \qw & \qw & \qw & \qw & \gate{\Rz(\frac{\pi}{2})} & \qw & \ustick{a_0} \qw & \qw & \qw & \qw & \qw & \qw & \qw & \qw& \ctrl{1} & \qw & \qw    & \qw & \qw & \qw & \ghost{\Sigma \frac{2}{3}} & \qw &  &  &  &  &  &
    \\ 
    \lstick{\ket{0}_{anc_0} }& \qw & \qw & \qw & \qw & \qw & \qw & \qw & \qw & \qw & \qw & \qw & \qw & \qw & \qw & \qw & \qw & \qw & \qw & \qw & \qw & \qw & \qw & \qw & \qw & \qw & \qw & \qw & \qw & \qw & \qw & \qw & \qw & \qw & \qw & \qw & \qw & \gate{H} & \control  \qw & \qw   
    \\ 
    \lstick{\ket{0}_{anc_1} }& \qw & \qw & \qw & \qw & \qw & \qw & \qw & \qw & \qw & \qw & \qw & \qw & \qw & \qw & \qw & \qw & \qw & \qw & \qw & \qw & \qw & \qw & \qw & \qw & \qw & \qw & \qw & \qw & \qw & \qw & \qw & \qw & \qw & \qw & \qw & \qw & \gate{H} &  \qw &  \qw & \control \qw & \qw 
    \\ 
    \lstick{\ket{0}_{anc_2}} & \qw & \qw & \qw & \qw & \qw & \qw & \qw & \qw & \qw & \qw & \qw & \qw & \qw & \qw & \qw & \qw & \qw & \qw & \qw & \qw & \qw & \qw &  \qw & \qw & \qw & \qw & \qw & \qw & \qw & \qw & \qw & \qw & \qw & \qw & \qw & \qw & \qw & \qw & \qw & \qw & \qw & \qw & \qw & \qw & \qw & \gate{H} &  \control \qw  & \qw & & & & & &
    \\ 
   & ~~\text{Cycle}  & & \text{1}& & & \text{2} & & & & & \text{3} & & & & & & & & & & & & & \text{4} & & & & & & & & & & & \text{5+6} & & & & \text{7} & & & & & \text{8+9} & & \text{10} & & & \text{11+12} & &  & 
    \\ 
    }} 
  \caption{\justifying\small Low-depth, highly-parallel circuit for computing the polynomial $P_6(x)=\frac{1}{7}(a_6x^6+a_5x^5+a_4x^4+a_3x^3+a_2x^2+a_x+a_0)$.}
    \label{fig:eh-poly_6_2}
\end{circuit*}

\section{Quantum Signal Processing (QSP)}
\label{secSM:QSP}

The QSP-type quantum circuit~\cite{Martyn2021GrandUnification} for computing a degree-$d$ polynomial $P_d(x)$ for a single real input can be constructed using a single qubit and $2d+1$ one-qubit operations. It consists of $d+1$ signal processing operators
\begin{equation}
S(\phi_k) =
\begin{bmatrix}
e^{i\phi_k} & 0 \\
0 & e^{-i\phi_k}
\end{bmatrix}
= e^{i\phi_kZ},
\end{equation}
with $\phi_k$ the angles encoding the desired polynomial, interspersed between $d$ signal rotation operators
\begin{equation}
W(x) =
\begin{bmatrix}
x & i\sqrt{1-x^2} \\
i\sqrt{1-x^2} & x
\end{bmatrix},
\end{equation}
which encode the input $x$.
The QSP unitary $U_{\phi}(x)$ can be represented as
\begin{equation}
U_{\phi}(x) =
\begin{bmatrix}
P_d(x) & \cdot\ \\
\cdot & \cdot\ \\
\end{bmatrix},
\end{equation}
where the real-valued target $P_d(x)$ is the top-left element of the unitary. The equivalent quantum circuit is shown in \cref{fig:qsp-seq1}.

\begin{circuit*}[htbp]
\centering
$\begin{myqcircuit}
    \lstick{|0\rangle}  & \gate{S(\phi_d)} &\qw & \cdots & &\gate{W(x)} & \gate{S(\phi_2)} & \gate{W(x)} & \gate{S(\phi_1)} & \gate{W(x)} & \gate{S(\phi_0)} & \qw \\
\end{myqcircuit}$
\caption{Quantum Signal Processing (QSP) sequence unitary $U_\phi(x)$ for fixed parity $P_d(x)$.}
\label{fig:qsp-seq1}
\end{circuit*}

However, to measure the value of $P_d(x)$, we need to perform a Hadamard test, which requires a second qubit and changes all $2d+1$ one-qubit operations to $2d+1$ controlled-angle rotations, resulting in the circuit in \cref{fig:hadamard-test}.

\begin{circuit*}[htbp]
\centering
$\begin{myqcircuit}
    \lstick{|0\rangle} & \qw & \gate{H} & \ctrl{1} & \gate{H} & \qw & \meter \\
    \lstick{|0\rangle} & \qw & \qw & \gate{U_\phi(x)} & \qw & \qw & \qw \\
    & & & U_\phi^H(x) & &
    {\gategroup{1}{3}{3}{5}{1.2em}{.}}
\end{myqcircuit}$
\caption{Hadamard test $U_\phi^H(x)$ measures QSP polynomial value $P_d(x)$.}
\label{fig:hadamard-test}
\end{circuit*}

Still, $U_\phi^H(x)$ can only compute fixed-parity polynomials. For undefined-parity polynomials, one needs to use Linear Combination of Unitaries (LCU)~\cite{Chakraborty2024LCU}, which requires an additional qubit and another layer of controls, shown in \cref{fig:lcu}.

\begin{circuit*}[htbp]
\centering
$\begin{myqcircuit}
    \lstick{|0\rangle}  & \gate{H} &   \ctrlo{1}  &   \ctrl{1} &  \qw  \\
    \lstick{|0\rangle}  & \qw & \multigate{1}{U_{\phi,\text{even}}^H(x) } & \multigate{1}{U_{\phi,\text{odd}}^H(x) } &  \qw & \meter\\
    \lstick{|0\rangle}  & \qw & \ghost{U_{\phi,\text{even}}^H(x) } & \ghost{U_{\phi,\text{odd}}^H(x) } & \qw  \\
\end{myqcircuit}$
\caption{Linear Combination of Unitaries (LCU) for arbitrary parity $P_d(x)$.}
\label{fig:lcu}
\end{circuit*}

The transpilation of \cref{fig:lcu} to two-qubit entangling gates results in approximately $12d+6$ CNOT gates. Alternatively, we can use Generalized Quantum Signal Processing (GQPS)~\cite{Motlagh2024GQSP} to encode arbitrary-parity polynomials, which requires similar resources.

\end{document}